# Evolutionary and Structural Constraints Define a Mutation-Resistant Catalytic Core in E. coli Serine Hydroxy methyltransferase (SHMT)

Deeptanshu Pandey[1], Dwipanjan Sanyal[2], Vladimir N. Uversky[3], Daniel C. Zielinski[4], and Sourav Chowdhury[1*]

**Affiliations:**

[1]Department of Biological Sciences, Birla Institute of Technology and Science-Pilani, Hyderabad, India.
[2]Department of Chemistry and Chemical Biology, Cambridge, Massachusetts, Harvard University
[3]USF Health Byrd Alzheimer's Research Institute, Morsani College of Medicine, University of South Florida, Tampa, FL 33612, USA
[4]Department of Bioengineering, University of California, San Diego, La Jolla, USA

[*]**Corresponding Author:**

**Sourav Chowdhury**
Department of Biological Sciences,
Birla Institute of Technology and Science, BITS-Pilani Hyderabad Campus,
Hyderabad, Telangana, India, 500078
Email: sourav.chowdhury@hyderabad.bits-pilani.ac.in

**Keywords:** Serine hydroxymethyltransferase (SHMT), Hamming distance, Shannon entropy, mutual information, intrinsic disorder, graph-based analysis, fitness trajectory, evolutionary analysis, protein conservation, *Escherichia coli*.

**Abstract**

Serine hydroxymethyltransferase (SHMT) is an essential enzyme in the *Escherichia coli* folate pathway, yet it has not been adopted as an antibacterial target, unlike DHFR, DHPS, or thymidylate synthase. To investigate this discrepancy, we applied a multi-scale computational framework that integrates large-scale sequence analysis of ~1000 homologs, coevolutionary interaction mapping, structural community analysis, intrinsic disorder profiling, and adaptive fitness modelling. These analyses converge on a single conclusion: the catalytic core of SHMT forms an exceptionally conserved and tightly coupled structural unit. This region exhibits dense coevolution, strong intramolecular connectivity, minimal disorder, and extremely low mutational tolerance. Peripheral loops and termini, in contrast, are far more flexible. Relative to established folate-pathway antibiotic targets, SHMT's active site is even more rigid and evolutionarily constrained. This extreme constraint may limit the emergence of resistance-compatible mutations, providing a plausible explanation for the absence of natural-product inhibitors. Fitness trajectory modelling supports this interpretation, showing that nearly all active-site residues tolerate only rare or neutral substitutions. Together, these findings identify SHMT as a structurally stable and evolutionarily restricted enzyme whose catalytic architecture is unusually protected. This makes SHMT an underexplored yet promising target for the rational design of next-generation antibacterial agents.

**Significance**

Serine hydroxymethyltransferase (SHMT) of *Escherichia coli* is a central enzyme in one-carbon metabolism but is not traditionally considered a drug target. Its evolutionary conservation and metabolic indispensability, however, make it an ideal model for understanding how essential enzymes balance catalytic precision with adaptive flexibility. In this study, we integrate graph-theoretic modular analysis, coevolutionary interaction networks, and Gaussian fitness landscape simulations derived from deep mutational data to dissect how SHMT's structural communities navigate mutational constraints. The analyses reveal a hierarchically organised architecture in which a rigid, epistatically constrained catalytic core is buffered by a peripheral layer of residues capable of supporting neutral or near-neutral transitions. This organisation enables the enzyme to maintain metabolic robustness while retaining limited potential for adaptive exploration under selective pressure. More broadly, this framework establishes a quantitative paradigm for investigating constrained evolvability in essential enzymes and highlights an emerging therapeutic perspective: that metabolic resilience can be perturbed not solely by targeting catalytic residues, but by disrupting the adaptive flexibility that underpins long-term enzyme stability and bacterial survival.

## Introduction

Serine hydroxymethyltransferase (SHMT) is a key enzyme in folate-dependent one-carbon metabolism, catalysing the reversible conversion of serine and tetrahydrofolate (THF) into glycine and 5,10-methylene-THF [1]. This reaction produces one-carbon units needed for nucleotide synthesis and methylation [2]. In *Escherichia coli*, SHMT is encoded by the glyA gene. Although glyA mutants can grow in rich media, they cannot survive in minimal media unless provided with serine or glycine, confirming the essential role of SHMT in metabolism [3]. E. coli SHMT serves as a strong model system for evolutionary and structural research because its structure has been solved at high resolution (e.g., PDB ID: 1DFO), its genetics are well understood, and it is highly conserved across bacteria [4,5]. As a result, we used E. coli SHMT as an experimentally manageable and broadly relevant system to investigate the structural constraints that influence its potential as an antibacterial target.

E. coli SHMT exists as a homo-tetramer, which is composed of two functional dimers, with each subunit binding the pyridoxal 5′-phosphate (PLP) cofactor [6] and folate substrate. The PLP cofactor forms a Schiff base with an active-site lysine to mediate one-carbon transfer, while stabilizing interactions, including π--π stacking (H126), salt bridges (D200), van der Waals contacts (A202), and hydrogen bonds (Y235, S99, H203), maintain substrate orientation and catalytically competent conformations [6]. Unlike eukaryotic SHMTs, which are mostly dimers with regulatory N-terminal extensions and distinct inter-subunit interfaces, bacterial SHMTs have a tetrameric architecture, reflecting an evolutionary trade-off between catalytic efficiency in prokaryotes and regulatory complexity in eukaryotes [7].

Despite its central role in bacterial one-carbon metabolism, serine hydroxymethyltransferase (SHMT) has been little explored as an antibacterial target [8]. Unlike dihydrofolate reductase (DHFR), which has multiple inhibitors but frequent resistance mutations, SHMT remains uncharacterised in this context. Its essentiality and deep evolutionary conservation raise the question of whether adaptive resistance through active-site mutations is feasible. The absence of SHMT-specific agents is striking given its essentiality and conservation across bacteria. However, the possibility of adaptive resistance through active-site mutations remains a concern [9]. Understanding SHMT's structural architecture, active-site topology, and evolutionary constraints is therefore critical for identifying druggable pockets and predicting mutational trajectories that may compromise long-term therapeutic efficacy, ultimately guiding the design of evolution-resistant inhibitors [9].

Although no SHMT-specific inhibitors are clinically approved, several small molecules inhibit the enzyme in vitro, including aminomethylphosphonate (AMP) analogues, hydroxypyrazoles, and oxalylglycine derivatives that mimic serine substrate, as well as classical antifolates such as methotrexate and 5-formyltetrahydrofolate [10]. These initial scaffolds highlight SHMT's susceptibility to small-molecule inhibition. Yet, unlike DHFR and DHPS, well-validated antibacterial targets, SHMT remains underexplored, likely due to historical focus on natural product scaffolds rather than lack of druggability [8]. Therefore, SHMT represents a promising but neglected antibacterial target.

We propose an integrative, multi-scale computational framework to explore the evolutionary and structural constraints of *E. coli* SHMT. We combined sequence conservation, co-evolutionary coupling, structural dynamics, and mutational fitness landscape modelling to dissect how this enzyme balances functional rigidity with limited adaptive flexibility. Beginning with a multiple sequence alignment of 999 homologues, we identified conserved residue positions constrained by strong evolutionary pressure and variable regions subject to greater mutational freedom. Co-evolutionary analysis revealed interdependent residue pairs, which, when mapped onto the protein structure, formed densely connected networks reflecting compensatory relationships that maintain catalytic function. Dynamic correlation analysis on the *E. coli* SHMT structure further revealed tightly coordinated macro-cores that are evolutionarily constrained. To capture this organisation, we applied structure-network analyses that result in structural communities, i.e., clusters of residues that interact, fluctuate and function in a coordinated manner. Within SHMT, these communities clearly separate a rigid, conserved catalytic core from peripheral regions with greater mutational tolerance, highlighting the modularity of its functional architecture. Our *in silico* highlights SHMT to be interpreted not merely as a set of conserved residues but as an integrated network architecture defined by evolutionary and structural interdependencies.

Finally, by modelling adaptive fitness trajectories of active-site residues, we assessed how evolutionary rigidity and tolerance to mutations are distributed within the catalytic site. Most positions showed highly resistant to mutation, with only a few residues capable of accommodating neutral substitutions without loss of function. This layered constraint architecture indicates that SHMT’s catalytic efficiency is evolutionarily preserved, with only limited flexibility at peripheral junctions. For antimicrobial design, this distinction is critical because inhibitors that target the conserved catalytic core are less likely to be compromised by resistance, while knowledge of the few mutationally susceptible sites helps in predicting possible escape routes. Therefore, SHMT emerges as an antibacterial target with strong resistance barriers, and our framework provides a broadly applicable strategy for identifying druggable sites in other essential enzymes, even in the absence of extensive mutational data.

## Results

The integrative workflow employed in this study combines large-scale sequence, structural, and evolutionary analyses to identify functionally and evolutionarily critical residues within a diverse set of SHMT**.** (Figure 1). We first generated sequence space from 999 SHMT sequences obtained from different species and analysed that by using Hamming distance matrices, PCA, and clustering to resolve global evolutionary relationships, positioning *E. coli* SHMT as a central scaffold within the homologous network. We studied multiple sequence properties/traits that identify conservation as well as co-variation of residue positions (Shannon entropy and mutual information, respectively), estimate their mutational tolerance and investigate intrinsic disorder. We integrated these features to define a catalytic core—a set of residues combining conservation, co-evolutionary coupling, and structural stability. Network and community detection further grouped these residues into broader macro-communities that capture two major types of constraints within SHMT: regions that are structurally essential for maintaining the enzyme's fold, and regions that are functionally critical for catalysis, substrate handling, or internal communication. Finally, adaptive walk and fitness trajectory modelling simulated mutational pathways to assess evolutionary accessibility and fitness optimisation. Collectively, this integrative pipeline prioritised residues and mechanistic insights that link evolutionary conservation, structural dynamics, and catalytic regulation across the SHMT family.

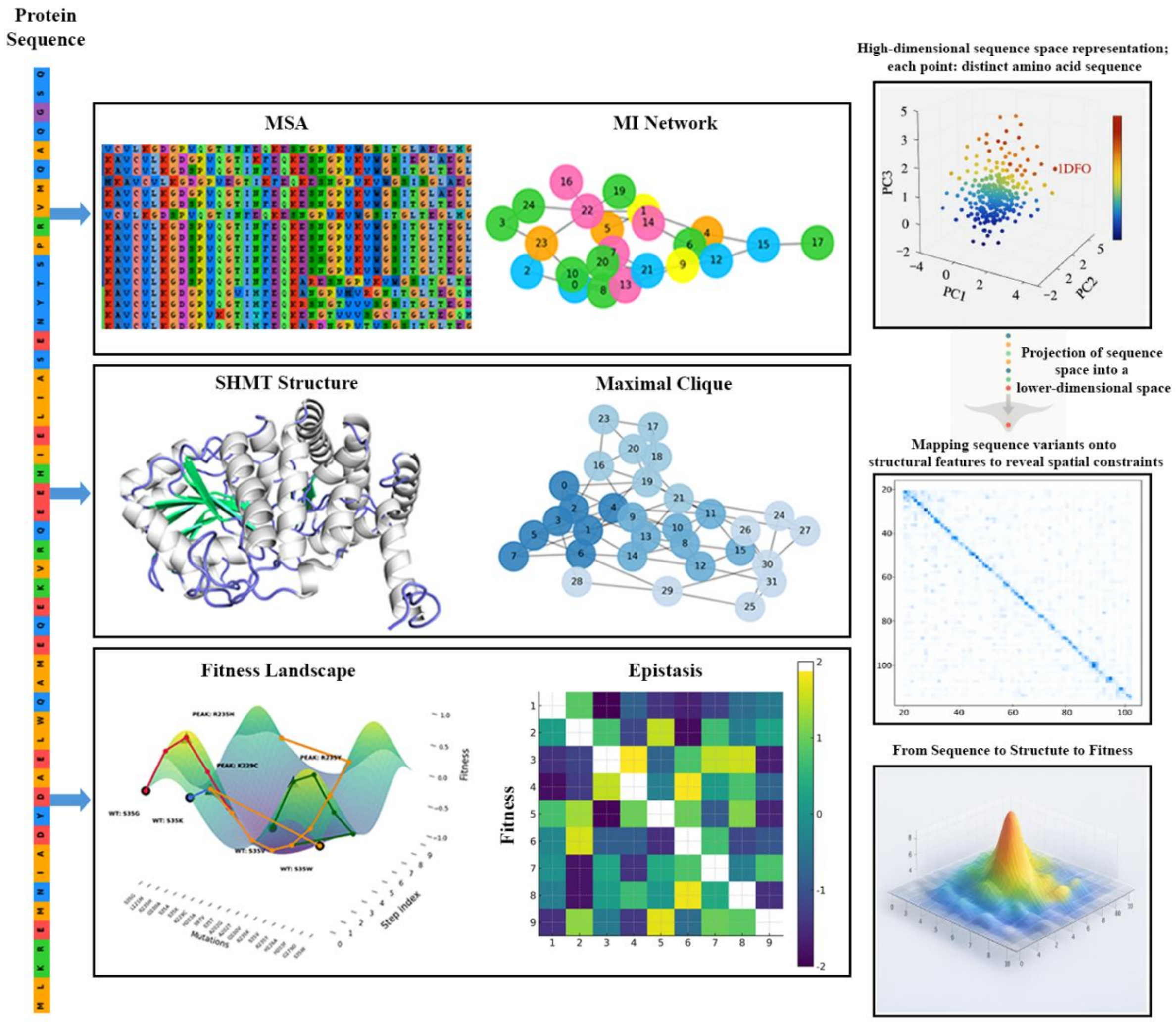

**Figure 1. Integrative mapping from protein sequence to structure to fitness landscape.** This framework links protein sequence variation to structural constraints and fitness outcomes through multiscale analyses. The workflow begins with the amino acid sequence (left), with residues color-coded for reference. Multiple sequence alignment (MSA) identifies conserved and variable positions, providing evolutionary context, while mutual information (MI) networks highlight co-evolving residue pairs and cluster them into communities. High-dimensional sequence space is visualized via PCA, colored by a fitness proxy, and a dimensionality reduction funnel illustrates projection into lower-dimensional space to reveal sequence constraints. Sequence variation is mapped onto the 3D SHMT structure, highlighting functionally important residues, and maximal cliques derived from the DCCM matrix identify highly interconnected co-evolving residue clusters. Contact maps reveal spatial proximities and structural constraints shaping accessible fitness peaks. Fitness landscapes capture the effects of single and combinatorial mutations, and epistasis matrices quantify synergistic or antagonistic residue interactions. A final integrative surface combines sequence, structural, and epistatic information, illustrating how these layers collectively define the constrained adaptive landscape of SHMT.

**Multiscale Sequence Analysis Reveals Conserved Constraints in Divergent SHMT Homologs**

Our analysis of 999 homologous SHMT sequences revealed a conserved evolutionary pattern in which E. coli SHMT (PDB ID: 1DFO) functions as a central ancestral scaffold, from which diverse homologs diverge while preserving the core structural and functional interaction network. Despite sequences from diverse lineages, the dataset showed strong taxonomic bias toward Pseudomonas (formerly Proteobacteria), including E. coli and related bacteria, densely populating the sequence space around 1DFO with close homologs. Multiple sequence alignment (MSA) [11] and subsequent data analyses revealed limited global divergence from this skew yet preserved local evolutionary context [4, 20] and minimal variation at critical residues [12], suggesting functional/structural constraints on residues for essential catalytic roles across species. Among the 25 closest sequences to E. coli SHMT, we saw that they all cluster very tightly, which shows strong purifying selection and very little change at important positions. This basically means that some residues can't afford to mutate because they are either needed for the enzyme's activity (functional constraint) or for keeping the protein properly folded and stable (structural constraint). To clean up the dataset, we used 90% identity clustering and removed synthetic or poorly annotated sequences, keeping only full-length, biologically meaningful ones. This gave us a good balance of diversity and quality, and helped us clearly identify conserved sites and build a better evolutionary framework for studying conservation, co-evolution, and structural interactions in E. coli SHMT [13, 14].

After aligning and cleaning 999 SHMT protein sequences, we quantified pairwise differences using the Hamming distance [15], generating a dissimilarity matrix across the dataset. Because direct visualization was impractical at this scale, we applied Principal Component Analysis (PCA) [16], to reduce dimensionality while preserving most variation. PC1 and PC2 captured the largest sources of variation, enabling a 2D projection of all sequences (Figure 2A). Each point represents an SHMT sequence, with spatial proximity reflecting similarity. E. coli SHMT (1DFO) appeared near the distribution's edge, indicating distinct substitutions relative to the central cluster of homologs. To test the robustness of this positioning, we performed bootstrapped PCA [17], repeatedly resampling 90% of sequences. The structure of the plot remained consistent across runs, confirming the reliability of the sequence relationships. K-means clustering [18] of the PCA output revealed three well-defined groups, highlighting natural clusters in SHMT sequence space (Supplementary Figure 4). A parallel analysis with t-SNE [19] produced similar clustering, reinforcing that 1DFO, while distinct, shares conserved features with a specific subgroup of homologs. Together, these validations confirmed that the PCA-derived structure was not an artefact [16, 19].

The apparent uniformity in Fig. 2B is biologically meaningful. The heatmap showed strong similarity among the 25 closest E. coli SHMT homologs, with only 1DFO sequence identical to the query and others exhibiting non-zero Hamming distances. This tight clustering reflects strong purifying selection on SHMT, an essential enzyme in folate metabolism, which restricts substitutions and preserves its catalytic core. The heatmap highlighted this constraint, with a block of dark green values indicating very low divergence. Thus, these 25 sequences defined a highly conserved neighbourhood with shared structural and functional features, consistent with purifying selection maintaining SHMT's core architecture [3, 6, 20].

After identifying the 25 closest sequences to E. coli SHMT in the PCA space, we found that this subset formed a remarkably tight and conserved cluster, suggesting strong functional similarity and evolutionary stability [20]. By evolutionary stability, we referred to the observation that these closely related SHMT homologs exhibit minimal sequence divergence, implying that selective pressures have maintained their structure and function over time. This high level of conservation points to essential roles played by SHMT in cellular metabolism, where even minor changes may compromise its activity [21]. To further explore which regions of the protein are under such constraint, we calculated Shannon entropy [22] across all aligned sequences, allowing us to pinpoint residues that have remained highly conserved throughout evolution. This residue-wise entropy profile allowed us to quantify the degree of sequence variability at each position [12]. To classify residues as conserved or variable, we employed three complementary thresholding strategies. First, a statistical approach using the mean entropy value (0.09) and one standard deviation above the mean (0.32) helped highlight sites with unusually high variability [12]. Secondly, we applied a non-parametric method based on the 25th and 75th percentiles [23] to avoid assumptions about the distribution shape. Finally, we used fixed cutoffs commonly referenced in literature [24], with entropy values below 0.10 considered conserved and those above 0.40 considered variable. These thresholds, visualised together in Figure 2C and compared in Supplementary Figure 1, produced consistent results across methods. A core group of residues, including Asp35, Gly57, Ser121, Tyr235, and Glu363, emerged as highly conserved in all approaches, with entropy values well below 0.1. These functionally essential residues are primarily involved in PLP cofactor binding and catalytic activity [5, 6, 25]. In contrast, higher entropy values at positions like 156, 308, and 399 pointed to more flexible, likely surface-exposed regions that may be under relaxed selective pressure or involved in species-specific adaptations [26]. Together, these results demonstrated that entropy-based conservation analysis can effectively distinguish structurally and functionally constrained residues from more variable, adaptable regions of the SHMT protein.

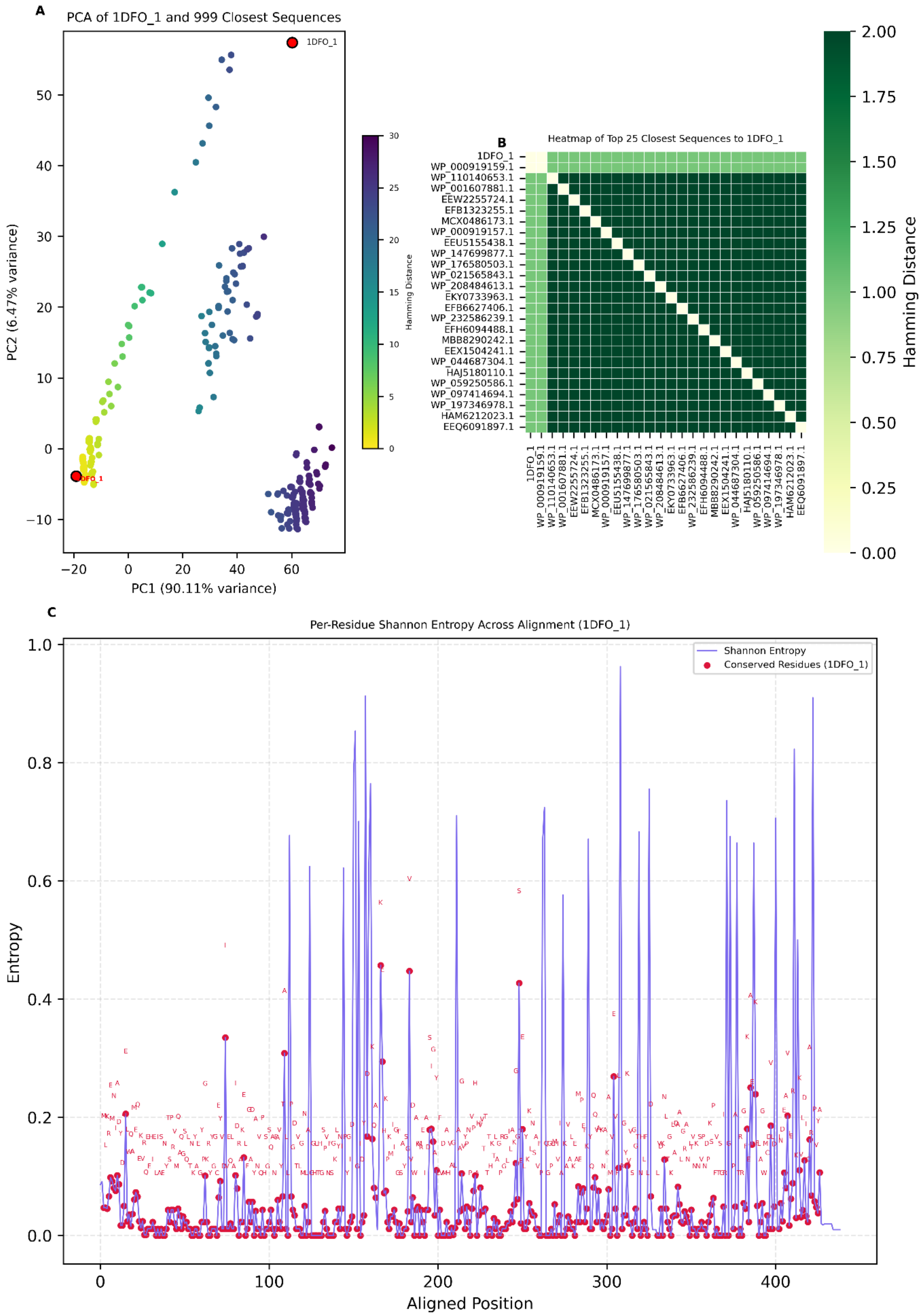
A
PCA of 1DFO_1 and 999 Closest Sequences
1DFO_1
PC1 (90.11% variance)
PC2 (6.47% variance)
Hamming Distance
B
Heatmap of Top 25 Closest Sequences to 1DFO_1
1DFO_1
WP_000919159.1
WP_110140653.1
WP_001607881.1
EEW2255724.1
EFB1323255.1
MCX0486173.1
WP_000919157.1
EEU5155438.1
WP_147699877.1
WP_176580503.1
WP_021565843.1
WP_208484613.1
EKY0733963.1
EFB6627406.1
WP_232586239.1
EFH6094488.1
MBB8290242.1
EEX1504241.1
WP_044687304.1
HAJ5180110.1
WP_059250586.1
WP_097414694.1
WP_197346978.1
HAM6212023.1
EEQ6091897.1
Hamming Distance
C
Per-Residue Shannon Entropy Across Alignment (1DFO_1)
Shannon Entropy
Conserved Residues (1DFO_1)
Entropy
Aligned Position

**Figure 2. Sequence Space Profiling and Evolutionary Variability of SHMT Using Homologs Closest to the Crystal Structure 1DFO: (A)** Principal Component Analysis (PCA) of 1DFO and its 999 closest homologous sequences derived from the Hamming distance matrix. Each point represents a single sequence and is coloured according to its Hamming distance from 1DFO. The position of 1DFO (red) lies at the centre of its closest mutational neighbourhood, reflecting strong evolutionary conservation among its nearest homologs. The first two principal components explain the majority of variation in the dataset (PC1: 90.11%, PC2: 6.47%), indicating that the dominant evolutionary differences among close SHMT homologs lie along a single major axis of sequence divergence, with secondary variation captured by PC2. (B) Pairwise Hamming Distance Heatmap for the top 25 most similar sequences to 1DFO. This matrix highlights the high similarity among close homologs, with a consistent dark green pattern reflecting minimal sequence divergence, supporting the use of these sequences for meaningful comparative analysis. (C) Per-residue Shannon entropy across the multiple sequence alignment of 1DFO with its 999 closest homologs. The entropy score on the y-axis indicates the degree of residue variability at each aligned position. Red circles mark the conserved residues of 1DFO, showing several low-entropy (high-conservation) stretches, especially in structured and functionally critical regions. Peaks in entropy identify positions with high sequence diversity, potentially corresponding to surface-exposed or non-critical residues. This evolutionary variability map provides insight into which residues are under selective constraint and helps frame the mutational tolerance analysis in subsequent figures.

**Coevolutionary Signals Reveal Structurally Coupled Residue Networks in SHMT**

While Shannon entropy highlighted residue-wise conservation in SHMT, it did not capture evolutionary dependencies between positions. Proteins function through coordinated conformational dynamics rather than isolated residues. Therefore, we extended our analysis using Mutual Information (MI) [27, 28]. MI quantified pairwise covariation across the alignment, identifying residues that tend to evolve together to preserve three-dimensional structure and biochemical function [29]. In contrast to entropy, which marks conserved sites, MI reveals inter-residue couplings that stabilise the catalytic core or maintain folding pathways. Together, these complementary approaches provided a more complete picture of how evolutionary pressures act not only on individual residues but also on cooperative networks within SHMT.

To assess residue interdependencies, we calculated MI scores for all residue pairs across the 999-sequence SHMT alignment [30]. We classified residue pairs with Z-scores above 3.0 and MI values greater than 1.0 as significantly coupled and used to build an MI interaction network (Fig. 3A) [31]. In this network, nodes represent residues and edges denote strong covariation, with node colours reflecting MI strength to emphasize clusters of coevolving sites. Residues with the highest MI values formed densely connected hubs, suggesting regions of strong evolutionary interdependence where substitutions are only tolerated through compensatory changes. Such hubs likely correspond to structural or catalytic centres critical for SHMT stability and activity. By contrast, residues with low MI scores (<0.4) were sparsely connected and tended to occupy peripheral positions in the network, consistent with greater mutational flexibility and weaker evolutionary coupling.

In parallel, we visualised MI scores in a matrix format (Supplementary Fig. 2). The heatmap revealed distinct high-scoring blocks, most notably between the N-terminal segment (Ser12 to Gly20) and the C-terminal region (Gly395 to Leu409). These long-range correlations suggest coevolutionary coupling between termini, potentially mediating interdomain communication or allosteric regulation [32]. Additional smaller patches of covariation were also observed in the protein core, including residues such as Phe156, Tyr174, and Ser235, consistent with modular coevolving units embedded within the structure. To relate these signals to SHMT's architecture, we mapped high-MI residues (MI > 1.0) onto the crystal structure (PDB ID: 1DFO; Fig. 3B) [4]. These residues were mainly located in the buried core of the protein, especially at helix–strand junctions and tightly packed interfaces. This suggests they help stabilise the protein's structure and support internal communication, meaning that movements or shifts in one part of the enzyme can propagate to other regions, allowing different parts to work together during catalysis. By contrast, low-MI residues were predominantly found on surface loops, flexible termini, or solvent-exposed regions, where mutational changes are more easily tolerated. This spatial distribution arises because high-MI residues overwhelmingly concentrate in the most structurally and functionally constrained regions of SHMT: 72% of all MI hotspots fall within 8 Å of the catalytic pocket and PLP-binding scaffold, and many occur at buried, low-RSA positions (RSA < 0.15, indicating deeply packed core residues that demand compensatory substitutions to maintain stability). In contrast, loop-exposed and dynamically flexible residues in macro-communities 3 and 4 exhibit high RSA values (RSA > 0.40, typical of solvent-exposed loops) and correspondingly low MI, indicating relaxed evolutionary pressure. Together, these trends show that coevolution hotspots align tightly with catalytic, cofactor-binding, and long-range communication residues, where geometric precision and structural coupling impose strong evolutionary constraints.

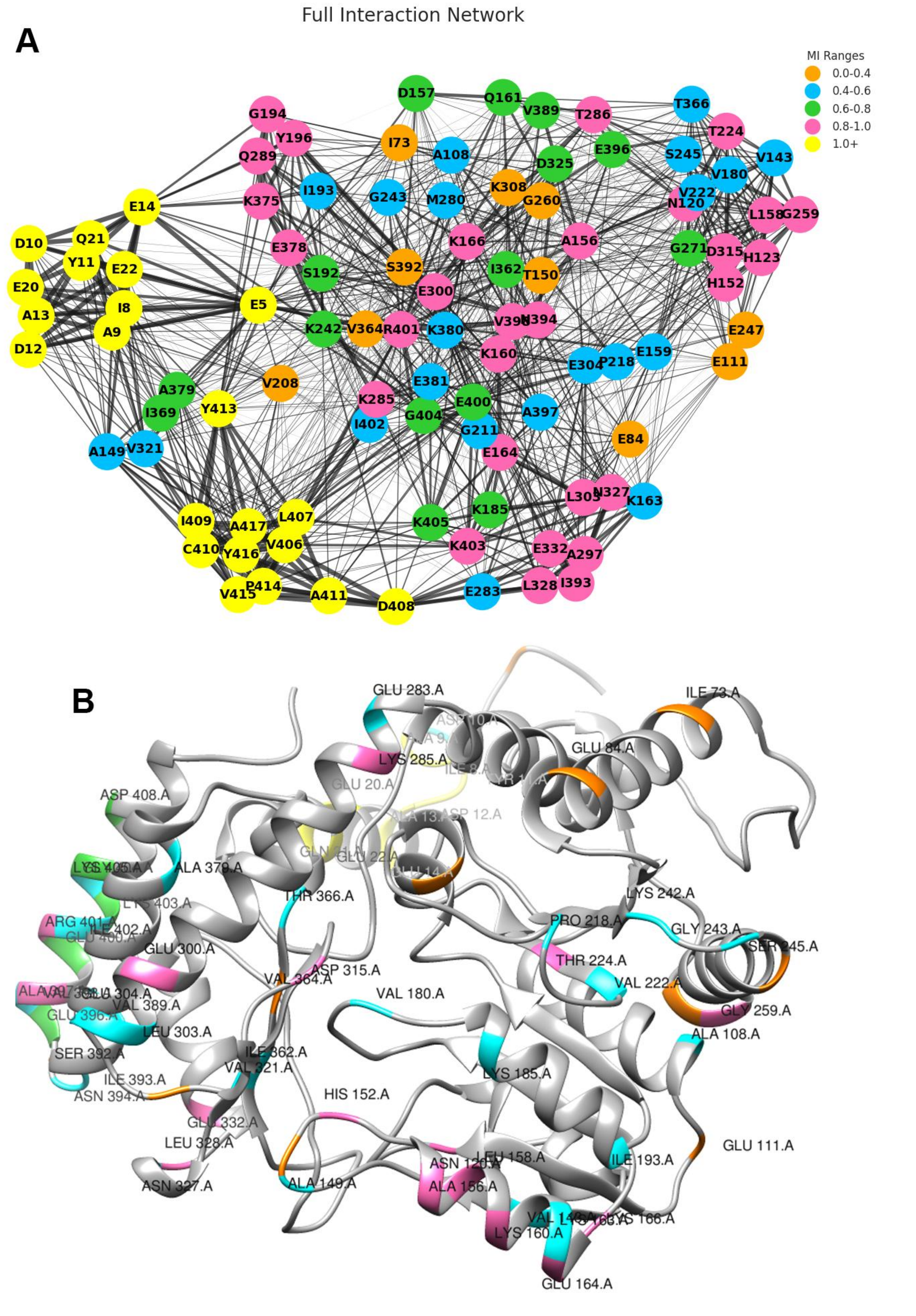

**Figure 3. Co-evolutionary Network Analysis Highlights Interconnected Residue Communities and Structural Coordination in SHMT:** This figure comprehensively visualizes co-evolutionary

residue interactions in SHMT, integrating network, structural, and community-level analyses. (A) A mutual information (MI)-based residue interaction network reveals dense connectivity, where nodes represent individual residues and edges represent co-evolutionary coupling above a given threshold. Residues are coloured based on their MI value, with yellow nodes (MI > 1.0) representing the strongest co-evolutionary links, often forming critical network hubs or connectors. (B) These co-evolving residues are mapped onto the SHMT 3D structure, showing that many high-MI residues are spatially clustered and bridge distant structural regions, implying potential roles in allosteric regulation or structural integrity

**Dynamic Cross-Correlation and Clique Decomposition Highlight SHMT's Functional Core**

To probe the dynamic organization of SHMT residues, we generated a residue interaction network from the Dynamic Cross-Correlation Matrix (DCCM) of backbone atomic fluctuations. In this network, nodes represent an amino acid residue (numbered according to the SHMT sequence), and edges connect pairs showing statistically significant correlated motions above the chosen threshold. This framework captures both long- and short-range dynamic couplings within the protein structure. We further identified maximal cliques, subsets of residues in which every member is directly connected to all others to define highly cooperative structural units. Distinctly colored cliques in the network revealed spatially contiguous, dynamically coordinated clusters, which likely contribute to catalytic efficiency, structural stabilization, and long-range allosteric communication (Supplementary Fig. 5).

Complementing the network analysis, a scatter plot of clique size versus average local clustering coefficient quantified the cohesiveness of these interaction modules (Fig 4A). Small cliques (2–5 residues) showed low clustering coefficients (0.01–0.05), consistent with weakly connected local interactions. Intermediate cliques (6–11 residues) showed moderate clustering (0.05–0.10), indicating partially cohesive interaction neighbourhoods, while larger cliques (≥12 residues) reached higher clustering values up to 0.14, highlighting highly cooperative residue interactions. This trend demonstrates that larger cliques correspond to tightly integrated, functionally and structurally significant modules, highlighting the hierarchical organization of SHMT dynamics where cooperative residue groups facilitate enzymatic function and structural integrity.

Building on this clique-level decomposition, we next applied a maximal clique-based community detection algorithm [33], which organized these tightly coupled cliques into 17 distinct residue communities (Fig 4B). Each community represents a dynamic unit of residues that move together, and many of them overlap with the high-MI hubs, indicating that coevolving residues are also embedded within larger-scale dynamical modules of the folded protein.

Among the detected groups, macro-community 2 emerged as the most densely interconnected and centrally located, comprising residues such as Ser53, His122, Glu155, Lys226, Tyr234, and Arg371. Many of which are highly conserved (low entropy), coevolving (high MI), and directly involved in PLP binding and catalysis (Fig 4C). We assessed functional enrichment in each macro-community by determining whether it contains a disproportionately high number of residues involved in catalysis, PLP–folate binding, structural stabilization, or long-range allosteric communication. Its strong network centrality and functional enrichment suggest that macro-community 2 represents the catalytic core of SHMT, integrating both enzymatic activity and structural stabilization. In contrast, macro-communities 3 and 4 were more peripheral and loosely connected, enriched in low-MI and dynamically flexible residues, typically located in loop regions or solvent-accessible surfaces, where evolutionary constraints are relatively relaxed. Macro-community 1 displayed intermediate characteristics, being more connected than the peripheral modules but likely functioning to support the global fold and facilitate allosteric communication. A complete list of the communities clustered into these macro-communities is provided in Supplementary Table 1.

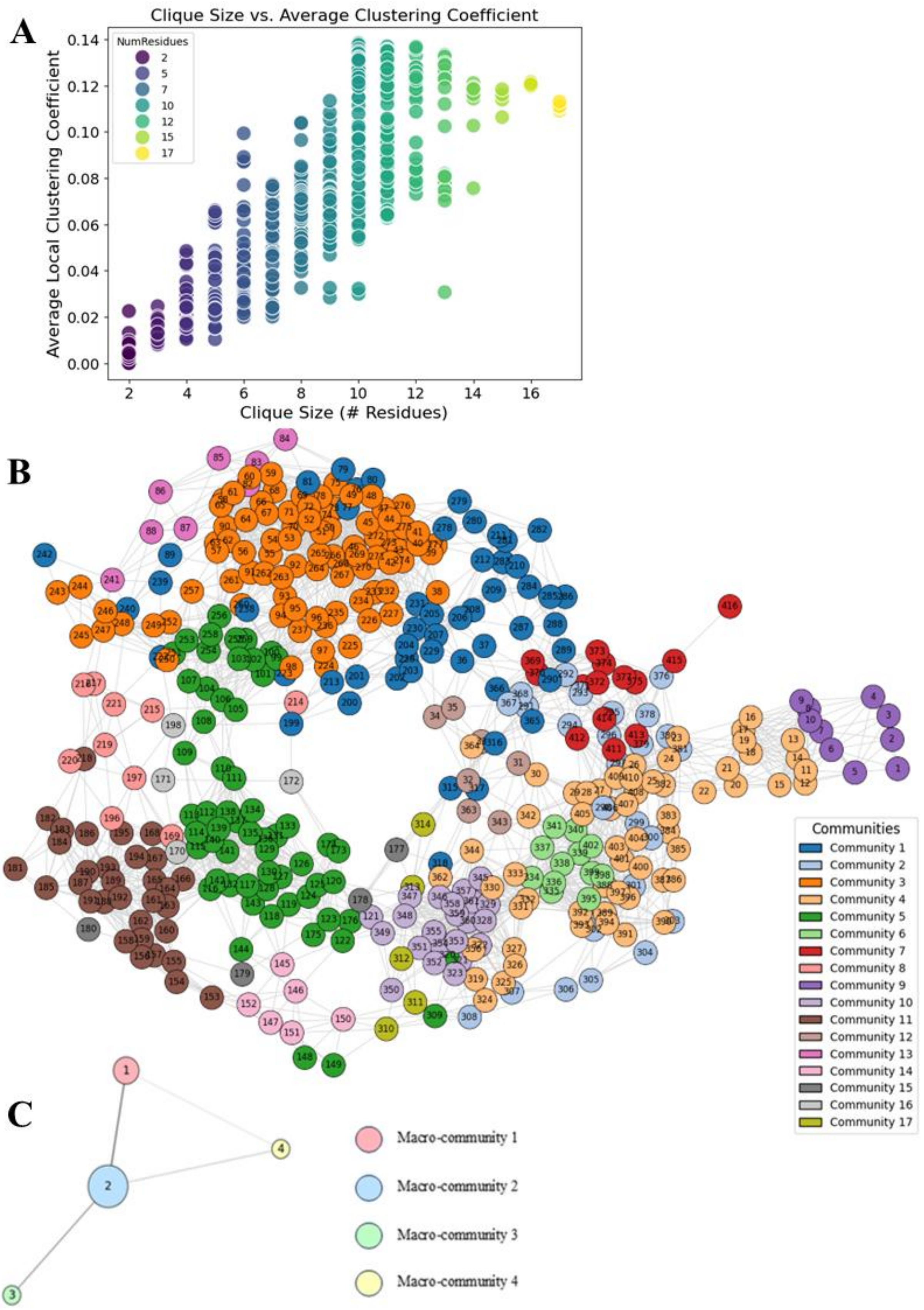

**Figure 4. Network analysis of SHMT dynamics and evolutionary constraints. (A)** Scatter plot showing the dependence of average local clustering coefficient on clique size derived from the Dynamic

Cross-Correlation Matrix (DCCM). Each point represents a maximal clique of residues, with the x-axis indicating clique size (number of residues) and the y-axis showing the average local clustering coefficient, a measure of local network cohesiveness. Points are color-coded according to clique membership (2–17 residues). Quantitatively, small cliques (size 2–5) display relatively low clustering coefficients in the range of 0.01–0.05, suggesting weakly connected local interactions. Intermediate cliques (size 6–11) show moderate clustering, typically between 0.05–0.10, indicating partially cohesive interaction neighborhoods. Larger cliques (≥12 residues) reach higher clustering values up to 0.14, highlighting strongly cooperative residue interactions and potential structural/functional modules within the SHMT residue network. **(B)** Network modularity analysis identifies 17 distinct communities of residues (each shown in a different colour), revealing groups of co-evolving residues that likely contribute to specific structural or functional subregions of the protein. **(C)** These 17 communities were further clustered into four macro-communities based on inter-community connectivity patterns, forming a simplified interaction graph where each node represents a macro-community and edge thickness reflects the strength of inter-macro-community interactions. Node size corresponds to the number of residues within each macro-community. This network analysis illustrates how evolutionary constraints organise SHMT into hierarchically structured residue communities that may underpin coordinated function

**Intrinsic Disorder Profiling Reveals Peripheral Flexibility and a Structurally Constrained Evolutionary Core in SHMT**

With the identification of evolutionarily constrained, co-evolving, and dynamically connected residues forming the structural and functional core of SHMT, we next investigated whether the remaining regions exhibit patterns of flexibility or structural plasticity. While conserved and tightly coupled residues are likely to support catalysis and stability, other segments might be more flexible and serve secondary roles. To investigate this aspect, we examined intrinsic disorder across the SHMT sequence, a feature often associated with conformational variability and specific biological roles such as allosteric regulation, signal transduction, molecular recognition, subcellular targeting, or serving as flexible linkers between structured domains [28, 34, 35]. These functions often rely on the ability of disordered regions to adopt multiple conformations or interact transiently with various partners.

To identify disordered segments within SHMT, we generated a residue-wise disorder profile using six independent predictors: PONDR® VLXT (black), PONDR® VSL2 (red), PONDR® VL3 (green), PONDR® FIT (pink), IUPred_short (yellow), and IUPred-Long (blue) [36-39]. Each algorithm relies on different biophysical principles to estimate disorder propensities, such as sequence complexity, hydrophobicity, charge distribution, and flexibility. By averaging their outputs, we created a mean disorder score for each position in the SHMT sequence, reducing the likelihood of predictor-specific bias and enhancing the reliability of the final profile.

The disorder plot (Figure 5A) showed that the majority of the SHMT sequence maintained disorder scores well below the threshold of 0.5, indicating a largely structured protein. However, several regions in the protein are predicted to have noticeable conformational flexibility, as only about 27% of the sequence is located below the threshold of 0.15. Furthermore, there are several prominent regions—residues 1–63, 81–106, 115–136, 150–171, 207–295, and 325–417—that consistently rise above the 0.15 cutoff and approach or exceed the 0.5 threshold across all the prediction methods. The use of a 0.5 threshold is biologically and statistically meaningful, as it is widely accepted in the literature to demarcate ordered versus disordered regions in proteins [40]. These regions represent prominent disorder "peaks" in the MDP (mean disorder prediction) profile, suggesting strong inter-method consensus and enhancing confidence in their identification as intrinsically disordered regions (IDRs) or flexible regions.

Each of the disordered segments corresponds to a specific region within the SHMT sequence. The first segment (residues 1–63) lies at the extreme N-terminus and forms the longest continuous stretch of predicted disorder. The second region (81–106) is a loop-like segment positioned between structured domains. The third segment (150–171) appears as a moderate peak and may act as a connector within the central part of the protein. The fourth region (207–295) forms a broader disordered stretch and is one of the widest peaks in the profile. Finally, the fifth segment (325–417) spans the C-terminal region and includes a second major disordered zone. Together, these segments frame the structured core of SHMT and are visually marked in Figure 3A as elevated peaks across all methods.

The presence of flexible regions in E. coli SHMT is further illustrated by Figure 5B, which represents a conformational ensemble generated for this protein by AFflecto [41]. Analysis of Figure 5B shows that high conformational flexibility is observed within the N- and C-terminal regions, as well as within the region with the widest disorder stretch (residues 207–295) found in intrinsic disorder analysis. These observations suggest that the well-packed 3D structure found for E. coli SHMT by X-ray crystallography is likely to originate from the PLP binding and dimerization of this protein.

To further understand how disordered regions relate to the protein's internal dynamics, we mapped the predicted disorder segments onto the macro-communities derived from DCCM-based analysis. This comparison showed that most of the disordered residues are located within macro-communities 3 and 4. These regions are situated toward the structural periphery of the protein and have fewer inter-community connections, suggesting they represent flexible domains that may support regulatory or adaptive functions rather than direct catalytic roles. In contrast, macro-community 2, previously identified as the most central and interconnected region, contains very few disordered residues. It is enriched in conserved, structurally ordered positions and overlaps with residues that show strong mutual information signals. This convergence of low disorder, high conservation, co-evolutionary coupling, and dynamic centrality strongly supports the idea that macro-community 2 forms the structurally stable and functionally critical core of SHMT.

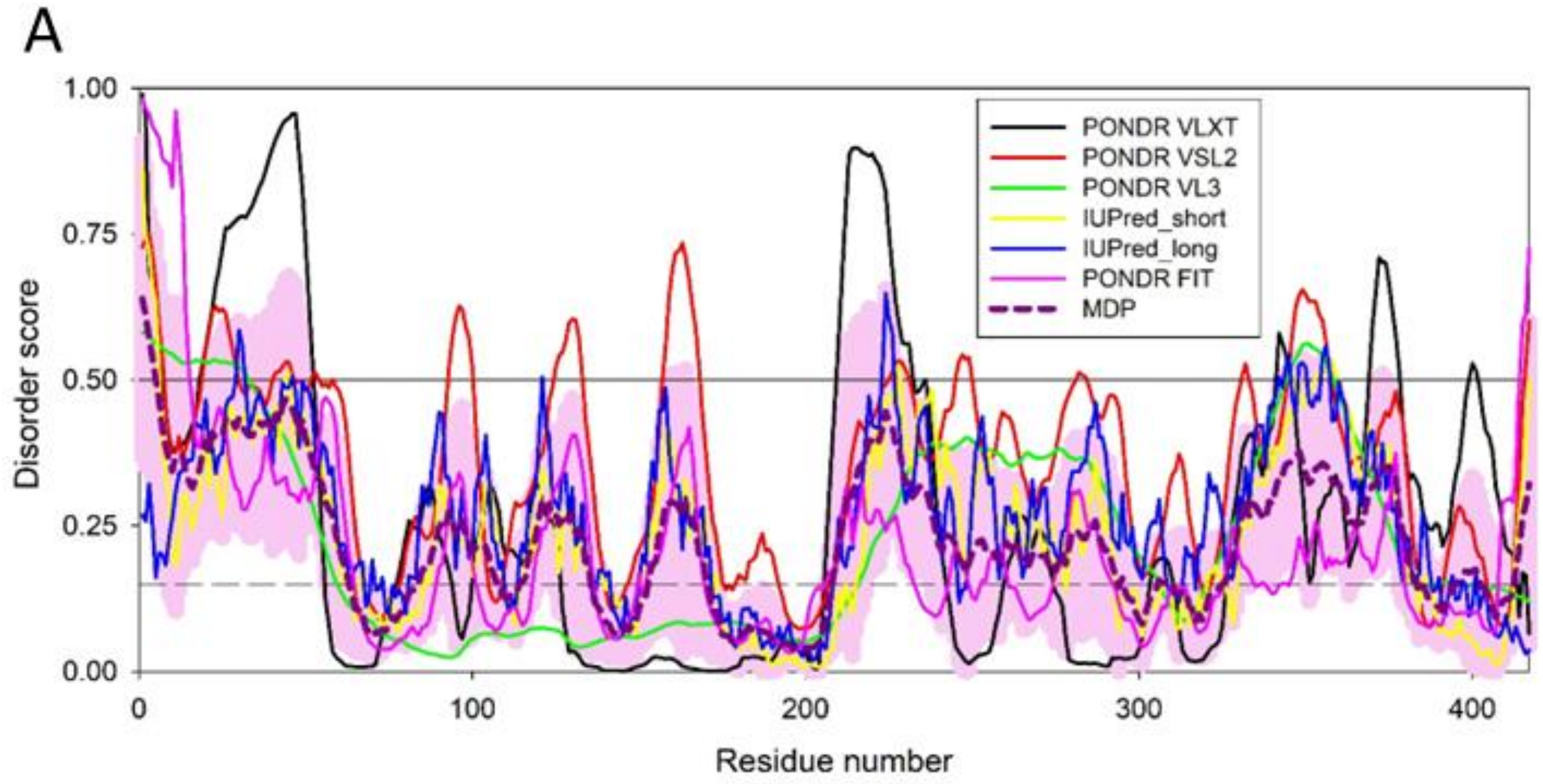

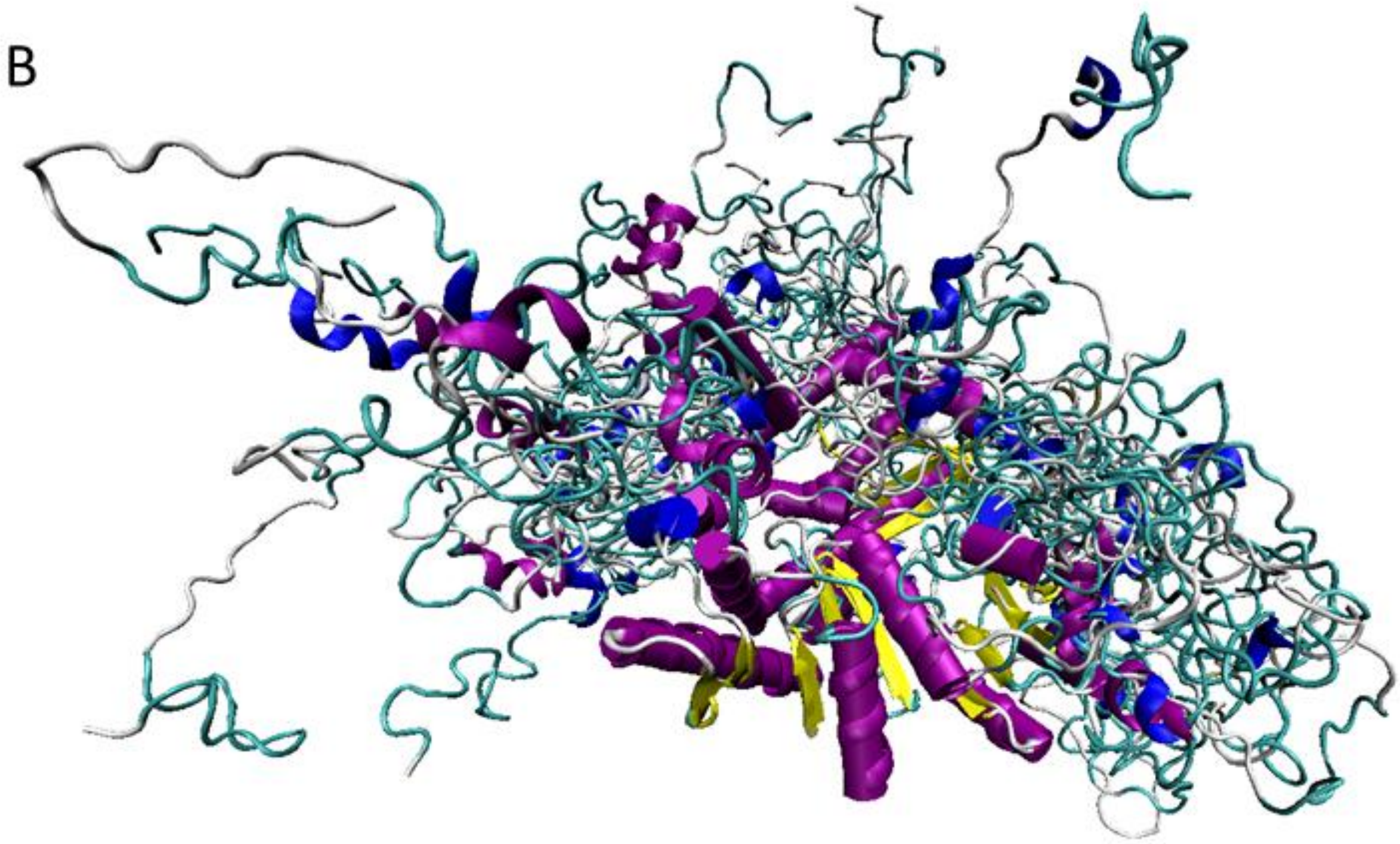

**FIGURE 5: Intrinsic Disorder Profiling Reveals Flexible Terminal and Loop Regions in SHMT (A)** Residue-specific disorder predictions for SHMT (UniProt ID: P0A825) were generated using five complementary algorithms: PONDR® VLXT (black), PONDR® VSL2 (red), PONDR® VL3 (green), PONDR® FIT (pink), IUPred_short (yellow), IUPred-Long (blue). The dark pink dashed line represents

the mean disorder score across methods, while the light pink shaded area denotes ±1 standard deviation, reflecting inter-method agreement. A black horizontal solid line marks the disorder threshold (0.5), above which residues are considered intrinsically disordered. A black horizontal dashed line shows the flexibility threshold (0.15). Data for this plot were assembled using Rapid Intrinsic Disorder Analysis Online (RIDAO) . Prominent disorder/flexibility-prone regions (corresponding to peaks in the disorder profile) include the N-terminus (1–63), inter-domain loops (~81–106, 115–136, 150–171, 207–295), and C-terminus (325–417), suggesting conformational flexibility in these segments. These findings align with prior Shannon entropy and MI analyses, linking structural disorder with mutational tolerance and evolutionary adaptability, while structured regions correspond to the conserved, co-evolving core. **(B)** Structural characterisation of *E. coli* SHMT (UniProt ID: P0A825) by AFflecto. A conformational ensemble consisting of 20 superimposed AFflecto-generated models is shown. α-Helices are depicted as purple helical coils; β-sheets are shown as flattened yellow arrows, where the arrow direction indicates the polypeptide chain N-to-C terminus orientation; coils or loops are represented as cyan irregular curves connecting the regular secondary structures. The original structure generated by AlphaFold is shown by cartoon, with α-helices being shown as purple helices. The plot was generated using Chimera software for molecular visualisation.

**Trajectory-Based Fitness Modelling Reveals Differential Adaptive Tolerance and Evolutionary Constraints Across SHMT Active-Site Residues**

After defining the modular and hierarchical organization of SHMT through entropy profiling, mutual information networks, dynamic community partitioning, and intrinsic disorder analysis, we next examined how residues particularly those belonging to the structurally conserved and functionally central Macro-community 2 navigate mutational constraints. In this framework, modularity represents the internal organization of SHMT into semi-independent but functionally coupled subunits that maintain dense intramodular interactions while communicating through defined interfacial nodes [42, 43]. These modules, identified through community detection and coevolutionary clustering, correspond to energetic and dynamic subsystems within the enzyme that together define its effective fitness landscape—a dynamic, population-dependent mapping of genotype to phenotype that captures the cumulative effects of mutation, selection, and constraint [44-46].

To probe how this network organization translates into evolutionary flexibility, we focused on six residues within Macro-community 2—S35, L121, H126, S175, D200, and R235—that exhibited high centrality in the residue-interaction network, low entropy, and strong coevolutionary coupling. Mapping these sites onto the E. coli SHMT crystal structure (PDB ID: 1DFO) revealed their clustering around the PLP-binding pocket and catalytic interface, forming a conserved core that integrates conformational dynamics and enzymatic control. Their physical proximity to the cofactor and substrate channels suggests that they act as regulatory pivots linking catalytic precision with conformational adaptability.

Building on this foundation, we performed adaptive-walk simulations using deep mutational scanning (DMS)–derived fitness predictions to model how sequential mutations alter local fitness. Each trajectory represents a hypothetical evolutionary route in which single amino-acid substitutions accumulate stepwise, and their cumulative effect on Z-scored relative fitness ($\Delta Z$) is recorded (Supplementary figure 6). The x-axis denotes the number of mutation steps, the y-axis represents relative fitness, and the color-coded surface shows the underlying effective fitness landscape. Crucially, neutral or near-neutral transitions ($\Delta$Fitness $< 0.05$) are now indicated by dashed line segments, marking trajectory intervals that impose minimal selective cost. Within the theoretical framework of neutral network theory, these dashed regions trace iso-functional genotype clusters that permit exploration of sequence space without loss of enzymatic competence [47-49]. They define the "corridors" of mutational neutrality that reconcile structural stability with evolutionary adaptability [50, 51].

The adaptive-walk trajectories (Figure 6) reveal distinct patterns of constraint and flexibility. Residues S35, S175, and D200 display steep, monotonic declines in fitness across successive substitutions, with no intervals signifying the absence of neutral transitions. These residues occupy highly rugged regions of the landscape and represent strongly constrained catalytic or structural sites. D200 (see Supplementary Fig. 7 for trajectories of residues D200)"., situated next to the PLP-binding cleft, forms an unfavourable acceptor–acceptor interaction with an adjacent carbonyl oxygen. Since two acceptors cannot form a stabilizing hydrogen bond, this interaction introduces a small amount of local electrostatic strain. Such strain restricts backbone flexibility, effectively 'locking' the PLP phosphate region in place and maintaining the strict geometric alignment needed for the enzyme's catalytic chemistry. S35, in the N-terminal loop, reinforces substrate positioning; and S175, within the interdomain hinge, coordinates conformational coupling. Mutations at these positions therefore impose immediate fitness penalties, consistent with evolutionary entrenchment residues fixed by selection to preserve catalytic precision.

In contrast, L121, H126, and R235 exhibit more gradual and variable fitness responses, each containing one or more discontinuous trajectory segments that denote neutral or near-neutral transitions. These residues occupy smoother, more navigable regions of the landscape, reflecting localized mutational buffering. L121 and H126, likely situated on solvent-exposed helices, tolerate limited side-chain variation while maintaining structural integrity. The most prominent case, R235, supports multiple extended dashed segments across independent trajectories, signifying consecutive neutral transitions and defining a locally buffered fitness basin. Such regions correspond to small-scale neutral networks that enable drift through iso-functional intermediates—an evolutionary strategy that preserves function while maintaining flexibility.

Collectively, these simulations reveal a layered adaptive architecture within SHMT's active site. A rigid catalytic nucleus (S35, S175, D200) is surrounded by partially flexible residues (L121, H126, R235) that provide limited mutational tolerance. This distribution represents the principle of hierarchical ruggedness, where deeply conserved cores are encased by partially neutralizable shells [52, 53]. These buffered regions occur at structurally permissive sites that allow small mutational steps without compromising the overall fold. Such duality rigid core with flexible periphery represents an optimized evolutionary design balancing robustness and evolvability.

Under effective fitness and dynamic landscape theory, SHMT shows a pattern in which core catalytic residues remain highly conserved, while peripheral sites allow limited but tolerable mutational change. Its landscape is not a single peak but a set of connected high-fitness paths and neutral steps that support constrained, stable evolution. This organization preserves catalytic function while providing just enough adaptive flexibility for long-term evolutionary maintenance.

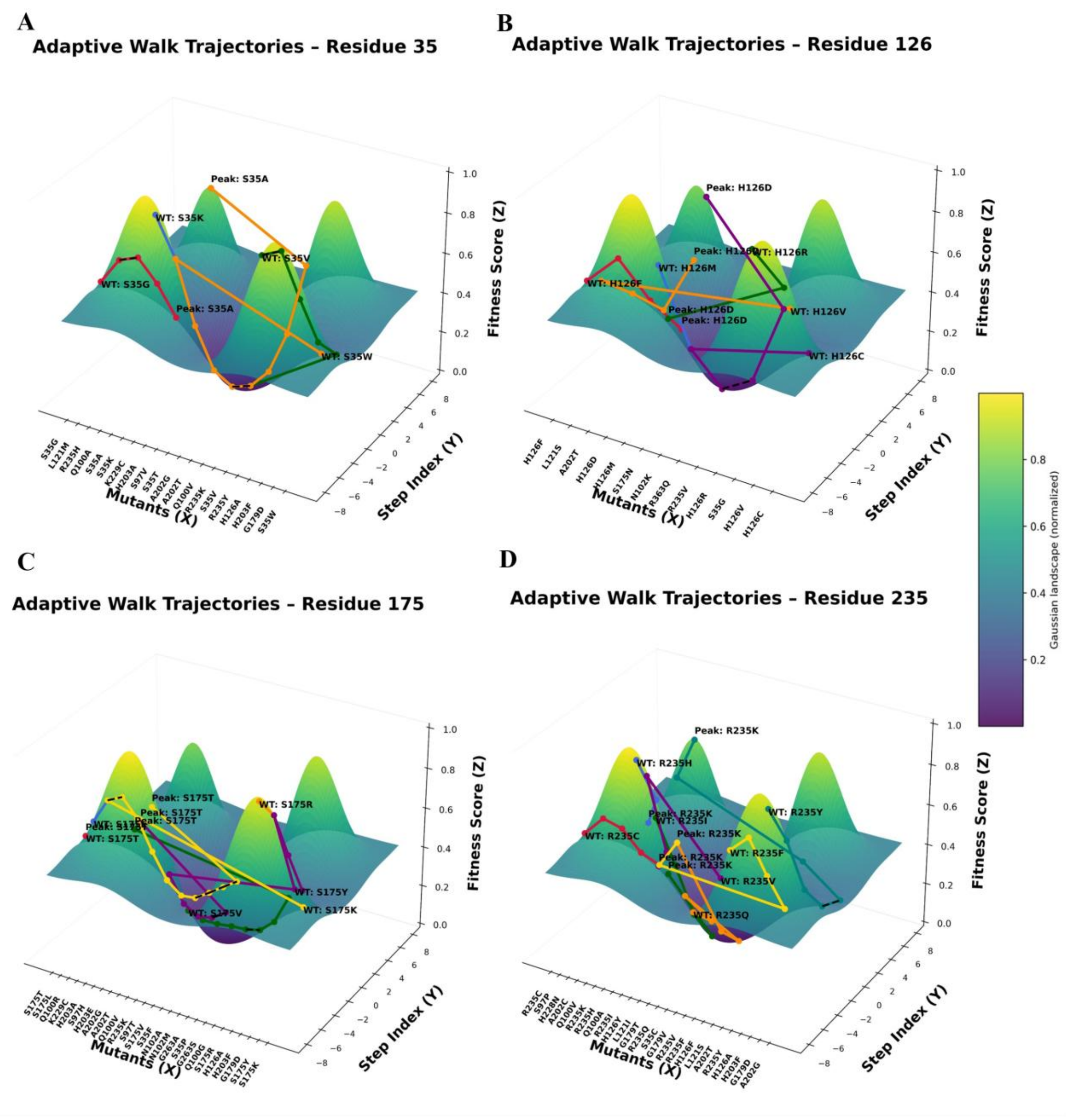

Figure 6. Adaptive-walk trajectories for four representative SHMT residues, derived from epistasis-informed fitness landscape simulations: Each panel presents a residue-specific fitness surface generated from epistatic interaction predictions, with adaptive-walk paths superimposed. The x-axis lists all possible single substitutions, the y-axis captures step progression along the walk, and the z-axis shows normalized fitness. Wild-type states, local optima, and neutral transitions (Fitness $< 0.05$) are annotated to highlight selectively accessible routes. These landscapes illustrate how epistasis shapes mutational accessibility and evolutionary dynamics across distinct local sequence contexts. (A) Residue 35. The landscape features broad gradients and multiple low-impact substitutions, indicating high mutational tolerance. Frequent neutral-accessible steps suggest weak local epistasis, enabling smooth and largely continuous evolutionary movement. (B) Residue 126. This landscape is markedly more rugged, with steep peaks and few neutral intermediates. The repeated convergence on H126D and H126R indicates

that only a narrow mutational subset maintains functional compatibility, reflecting strong epistatic constraint. (C) Residue 175. The topology alternates between distinct fitness basins separated by occasional neutral steps, consistent with a partially buffered site. A limited set of permissive mutations allows movement between basins, enabling both gradual and jump-like adaptive transitions.(D) Residue 235. A single dominant peak (R235H) is surrounded by sharply deleterious substitutions, with no neutral plateaus. This strongly canalized landscape indicates that residue 235 is a critical structural or functional node with minimal evolutionary flexibility.

## Discussion

The folate pathway is a central metabolic route in Escherichia coli (E. coli), responsible for producing cofactors critical for nucleotide and amino acid biosynthesis [1, 2]. Within this pathway, serine hydroxymethyltransferase (SHMT) catalyzes the reversible conversion of serine and tetrahydrofolate into glycine and 5,10-methylene tetrahydrofolate [1, 3]. This reaction forms a key node in folate-mediated one-carbon metabolism, supplying one-carbon units essential for thymidylate, purine, and methionine biosynthesis, thereby linking amino acid metabolism with nucleotide synthesis and overall cellular growth. Given its critical role, SHMT has emerged as an attractive target for antimicrobial drug development, especially for treating bacterial infections. Unlike well-studied targets such as dihydrofolate reductase (DHFR), which has numerous documented resistance mutations [61, 62], E. coli SHMT has no known resistance-linked variants. This likely reflects its status as a nontraditional drug target rather than an inherent inability of the enzyme to acquire adaptive mutations. It suggests that SHMT may be evolutionarily constrained, potentially making it less tolerant to mutations that could compromise its catalytic efficiency (Kcat). This functional pressure could limit the potential for resistance, making SHMT an even more attractive drug target for future therapies. Understanding these structural and evolutionary constraints could provide valuable insights for developing more effective SHMT-targeting drugs with a lower risk of resistance emergence.

Mutations in these residues can profoundly influence turnover rate, substrate preference, and overall stability, thereby constraining SHMT's evolutionary flexibility. Nonetheless, certain conservative or near-neutral substitutions such as polar-to-polar or small hydrophobic-to-hydrophobic replacements around the active site may be tolerated without major loss of function. These subtle alterations could preserve catalytic performance while gradually expanding the enzyme's adaptive landscape, potentially serving as precursors to resistance under sustained selective pressure [55, 60]. The absence of comparable mutational data for SHMT thus represents a major knowledge gap, emphasising the need for comprehensive deep mutational scanning to reveal how specific active-site changes shape both catalytic efficiency and evolutionary resilience in drug-challenged environments [20, 47].

Our multiscale analyses revealed a dual-layered evolutionary architecture in SHMT. Although E. coli SHMT appears divergent within the broader global sequence space, it forms a tight cluster with a subset of highly conserved homologs [11, 13]. This pattern reflects the principle of modular protein evolution, wherein distinct structural regions evolve at different rates [42, 49]. The catalytic core remains under strong purifying selection, preserving residues essential for enzymatic function, while peripheral and regulatory regions exhibit higher evolutionary plasticity allowing adaptation to species-specific requirements such as allosteric regulation, metabolic context, or interactions with partner proteins [40, 49].

Shannon entropy highlights residues that are evolutionarily conserved [22, 24], yet proteins function as integrated systems in which the role of each amino acid depends on its interactions with others [30]. To capture these interdependencies, we employed mutual information (MI) analysis to identify co-evolving residue pairs [31, 32], revealing statistically coupled networks suggestive of compensatory or cooperative dynamics. These co-evolving residues were distributed non-randomly, clustering within SHMT's structural core and marking regions of strong functional interdependence. The emergence of MI-dense modules implies that structural or functional changes are often tolerated only through coordinated substitutions, reflecting a modular organization that preserves internal stability. When mapped onto SHMT's three-dimensional structure, high-MI residues were found concentrated within buried regions at α-helix and β-strand interfaces sites critical for maintaining folding stability and allosteric communication [28, 70]. These interconnected clusters resemble "evolutionary sectors" or "allosteric cores," representing subnetworks of residues that evolve cooperatively to sustain SHMT's sub-structural balance and dynamic functional integrity [63]. To assess whether these modules also exhibit dynamic coherence, we applied a dynamic cross-correlation matrix (DCCM) approach [57, 70]. The resulting networks showed tightly linked residue communities with coordinated motion, translating evolutionary coupling into mechanical communication. A strong overlap between MI hotspots and DCCM communities supports a modular, systems-level view of SHMT, where evolutionary constraints and structural dynamics converge to shape functional coordination and mutation response.

The hierarchical organization of SHMT's residue interaction network revealed four distinct macro-communities with clear spatial and functional demarcations within the monomer. Macro-community 2, which encompassed the majority of densely interconnected residues from Communities 1–3, 5–8, 11, 15, and 17, occupied the central α6–β8 catalytic domain—the structural and functional core of the enzyme. This region harboured most of the PLP–folate binding and catalytically essential residues, including SER35, GLY98, SER99, ASN102, HIS126, ASP200, ALA202, HIS203, THR226, LYS229, ARG235, HIS228, and ARG363, spanning the β-sheet core and its surrounding α-helices that stabilise the cofactor-binding pocket. The high density of mutual information-defined residues and low-entropy positions within this region reflects its coevolutionary rigidity, suggesting that it acts as an optimised catalytic scaffold where even minor perturbations may propagate through multiple dynamic coupling pathways, strongly impacting enzymatic function. In contrast, macro-community 1, formed by communities 4, 10, 13, 14, and 16, was distributed across the N-terminal arm and peripheral α-helical regions (α1–α4, α12–α14), exhibiting higher conformational flexibility and lower MI density. This peripheral domain likely mediates species-specific adaptations, regulatory interactions, and allosteric communication with metabolic partners rather than direct catalysis. Macro-community 3 (Community 9) occupied a loop–helix junction adjacent to the C-terminal interface, representing a partially flexible connector that may facilitate structural buffering and domain motion between the catalytic and peripheral regions. Meanwhile, Macro-community 4 (Community 12) extended along the C-terminal tail and β9–β10 sheet cluster, forming a flexible but topologically cohesive segment that contributes to dimer stabilisation and long-range conformational coupling. Together, these macro-communities delineate SHMT's dual-layered evolutionary architecture a tightly coevolved catalytic hub embedded within a more adaptable periphery highlighting how evolutionary constraints and modular dynamics coalesce to preserve catalytic fidelity while permitting functional innovation in response to environmental and metabolic pressures.

Intrinsic disorder predictions further confirmed this core-periphery architecture [37, 38, 39, 40, 41]. Using an ensemble of disorder prediction algorithms, we identified five discrete regions with elevated disorder scores, all overlapping spatially with surface-exposed loops or terminal segments. These intrinsically disordered regions (IDRs) displayed high sequence entropy and minimal MI, suggesting that they evolve independently and are structurally decoupled from the catalytic core [35]. Functionally, these disordered elements may act as flexible linkers, facilitate protein–protein interactions, or undergo disorder-to-order transitions upon ligand binding [34, 41].

The influence of this architectural modularity on SHMT's mutational adaptability became especially apparent through trajectory-based modelling of active-site residues [53, 54, 60]. By simulating mutational paths across the fitness landscape, we observed that some residues, such as 102, 203, 228, and notably 363, could accommodate neutral mutations, suggesting local buffering mechanisms or redundancy in functional contributions. These residues are likely situated at conformationally flexible junctions or non-catalytic yet functionally adjacent regions of the active site. Their ability to tolerate mutations without loss of function implies potential latent pathways for adaptation under selective pressure. On the other hand, residues such as 100 and 202 displayed strictly increasing fitness trajectories devoid of any neutral steps, indicating an extreme sensitivity to perturbation [54]. These correspond to residues directly involved in substrate binding or catalytic chemistry, where even minor structural alterations severely compromise Kcat.

These observations indicate that mutational constraint in SHMT is not uniformly distributed even within the active site but is organized into zones of varying rigidity. This partitioning mirrors principles observed in drug resistance evolution, where "hotspot" residues are highly constrained, while adjacent permissive regions can accumulate neutral or compensatory mutations [55, 61]. Well-studied examples include Plasmodium falciparum DHFR, where resistance-conferring mutations at key catalytic residues (e.g., N51I, C59R, S108N) are enabled by nearby permissive substitutions like I164L, which buffer fitness costs [61]. Similarly, in Mycobacterium tuberculosis KatG, the S315T resistance mutation is stabilised by distal compensatory changes [68].

Near-neutral mutation sites that do not drastically reduce catalytic activity often mark sensitive points in the enzyme's internal communication network. These residues typically sit at the boundaries between rigid catalytic regions and flexible peripheral domains, where small structural shifts can subtly influence how signals or motions spread through the protein. In SHMT, positions such as Arg363 and nearby loop–helix junctions illustrate this principle: although mutations here may not immediately disrupt catalysis, they can modulate the enzyme's conformational dynamics and responsiveness. Recognizing such positions helps guide allosteric modulator design, as small molecules targeting these flexible, dynamically connected regions can alter SHMT's activity indirectly—for instance, by changing how the enzyme fluctuates or how different domains move relative to one another. This strategy expands the range of possible inhibitor sites beyond the crowded active site and offers a lower risk of resistance, since these residues are evolutionarily tolerated but still tightly linked to essential catalytic behavior. In this way, near-neutral mutation sites serve as natural indicators of potential allosteric control points that can be harnessed in next-generation antimicrobial development.

Molecular dynamics (MD) simulations can help reveal how mutationally tolerant regions of SHMT behave over time [57]. Some residues may undergo small structural changes that keep the active site intact while allowing flexibility. Studying these dynamics can show how proteins tolerate mutations and, together with mutagenesis and functional assays, improve understanding of SHMT's resistance and adaptability.

Our work illustrates how SHMT's structural dynamics, evolutionary history, and catalytic requirements are closely interconnected, shaping the enzyme's architecture to balance stability, flexibility, and function. By combining co-evolutionary mapping [31, 32, 33], dynamic community detection [33, 57], disorder analysis [37, 38, 39, 40, 41], and trajectory-based modelling [53, 54, 60], we present a detailed view of SHMT's mutational landscape. These results show how patterns of rigidity and flexibility determine the enzyme's balance between stability and adaptability. Understanding this network of constraints offers a framework for developing new folate-pathway modulators or nonclassical antifolate analogs that target SHMT more precisely. Such approaches could lead to evolution-proof antibacterial strategies with lower risk of resistance and guide enzyme engineering for metabolic rewiring and synthetic biology applications.

**Methods**

Using keyword and domain-based searches tailored to serine hydroxymethyltransferase (SHMT), we first obtained a non-redundant dataset of 999 protein sequences from the UniProtKB database (release 2024_03) in order to evaluate evolutionary divergence among SHMT homologs. Sequences were filtered according to functional annotation, protein length consistency, and sequence completeness to guarantee precise homolog identification. To reduce the over-representation of closely related strains, redundancy reduction was carried out using CD-HIT v4.8.1 at a 90% identity threshold [18]. The final set was confirmed to span multiple bacterial phyla when the taxonomic annotation of the dataset was confirmed using NCBI Entrez (accessed via Biopython v1.81) [19]. However, the distribution was skewed toward Pseudomonadota (formerly Proteobacteria), reflecting the phylogenetic proximity to E. coli.

To assess the evolutionary divergence among SHMT homologs, a pairwise Hamming distance matrix was constructed using protein sequences aligned with MAFFT v7.505 [10]. Hamming distance is a direct and interpretable measure of sequence dissimilarity, defined as the total number of mismatched residues at corresponding positions in two aligned sequences [17]. For any sequences A and B of equal length L, the Hamming distance $D_h(A, B)$ is computed as the sum over all positions i of the indicator function $\delta(A_i, B_i)$, where $\delta = 1$ if $A_i \neq B_i$ and zero otherwise. That is, $D_h(A, B) = \Sigma (A_i \neq B_i)$, for i = 1 to L. The resulting matrix is symmetric and of size N×N times, where N is the number of homologs. A matrix heatmap was then generated to visualise sequence similarity patterns, with darker regions indicating low divergence and lighter regions indicating higher dissimilarity [22]. This visualisation was an initial step in identifying conserved subgroups and divergent clusters among the homologues.

Before performing dimensionality reduction, the distance matrix was standardised using z-score normalisation [19]. Standardisation ensures that each feature contributes equally to the analysis by removing scale-related bias. The normalised value z of a given feature x is calculated as $z = (x - \mu) / \sigma$, where $\mu$ is the mean and $\sigma$ is the standard deviation of the feature vector. This step was implemented using the Standard Scaler function from scikit-learn v1.2.2 [19]. Normalising the matrix was crucial for Principal Component Analysis (PCA) to correctly extract the principal axes of variation without being skewed by feature magnitudes [24].

Principal Component Analysis was then applied to the normalised distance matrix to reduce dimensionality. PCA transforms the original high-dimensional data into a lower-dimensional space while retaining the maximum variance possible. Each SHMT sequence was projected onto two principal components (PC1 and PC2), which accounted for the most significant variation in the dataset. The transformation of a standardized data point x into the reduced space is given by: $y = W^t (x - \mu)$, where W is the matrix of eigenvectors (principal component loadings), and $\mu$ is the mean vector of the data. The result was a two-dimensional representation of sequence relationships, where each point corresponded to a homolog positioned according to its overall sequence dissimilarity to others. K-means clustering was applied to the two-dimensional coordinates to identify natural groupings in the PCA-reduced space [20]. This algorithm partitions the data into k clusters by minimising the within-cluster variance. The K-means objective function is defined as:

$$J=\Sigma_k\Sigma_i||x_i-\mu_k||^2,$$

Where $x_i$ is a data point in cluster k, and $\mu_k$ is the centroid of that cluster. The optimal number of clusters, k, was determined using the elbow method and silhouette analysis [25]. The elbow method involves plotting the within-cluster sum of squares (WCSS) against various values of k and selecting the value at which the decrease in WCSS begins to level off. The silhouette coefficient evaluates how well each data point fits within its assigned cluster versus others, calculated as:

$$s(i) = (b(i) - a(i)) / max(a(i), b(i))$$

A (i) is the average intra-cluster distance, and b(i) is the average nearest-cluster distance. Both analyses indicated that three clusters (k = 3) provided the most coherent grouping of SHMT homologs. The final PCA coordinates and cluster assignments were visualised using Matplotlib v3.7.1 [19]. Each sequence was plotted in the PC1–PC2 plane and colour-coded according to its cluster membership. The reference Escherichia coli SHMT sequence (PDB ID: 1DFO) was specifically annotated and highlighted in red to examine its relative positioning within the homolog landscape [4].

This integrative approach, starting from Hamming distance quantification, followed by dimensionality reduction via PCA, and culminating in unsupervised clustering, enabled a clear, interpretable, and statistically supported classification of SHMT homologs into meaningful evolutionary subgroups.

*Shannon Entropy Calculation*

Shannon entropy (S(*i*)) was calculated to quantify the variability of amino acids at each sequence position. Positions with low entropy values represent conserved residues, critical for structural and functional integrity, while high entropy values indicate positions prone to mutations [23].

$$S(i) = -\sum_{a_i=1}^{20} P(a_i) \log P(a_i)$$

Where *i* represents the sequence position, $P(a_i)$ is the probability of amino acid *a* occurring at position *i* in the MSA. Lower values of S(i) correspond to highly conserved residues, while higher values suggest increased variability.

We used three statistically supported thresholding techniques to categorise residues according to evolutionary constraint. First, a parametric approach that defines moderate and high variability thresholds using the mean entropy and one standard deviation above it, a common procedure in statistical outlier and dispersion analysis [23]. Second, a non-parametric method that captures robust distributional cutoffs without assuming normality by utilising the 25th and 75th percentiles of the entropy distribution, frequently employed in exploratory data analysis [23]. Third, we employed fixed entropy thresholds, which are widely used in studies applying entropy metrics for protein conservation and functional site prediction (0.1 for conserved, 0.4 for variable) [24].

To evaluate the robustness of residue classification using these approaches, a sensitivity analysis was conducted. Based on their entropy values under each scheme, residues were classified as highly variable, moderately variable, or highly conserved. A comparative table and bar plots with thresholds superimposed were used to illustrate the findings. Regardless of any one statistical cutoff, this multi-threshold approach guaranteed the accurate and biologically significant identification of conserved residues.

*Mutual Information Calculation*

To investigate co-evolving residues, we calculated mutual information (MI) for each pair of positions in the MSA [33,34]. MI quantifies the degree of correlation between two sequence positions, revealing residues that might be functionally or structurally interdependent. These co-varying residues could indicate regions where mutational compensation occurs, maintaining overall protein function despite mutations in specific positions.

$$MI(i,j) = \sum_{a_i=1}^{21} \sum_{b_j=1}^{21} P(a_i, b_j) \log \frac{P(a_i, b_j)}{P(a_i)\, P(b_j)}$$

In this formula, $P(a_i, b_j)$ represents the probability of amino acids *a* and *b* occurring at positions *i* and *j*, respectively, while $P(a_i)$ and $P(b_j)$ are the individual probabilities at each position. The MI(i,j) values range from 0 (uncorrelated residues) to MImax (the most interdependent residue pairs). The MI values were visualised as a heatmap and a co-evolutionary matrix, generated using Python [35].

*Intrinsic Disorder Prediction*

Intrinsic disorder prediction was performed on *Escherichia coli* Serine Hydroxymethyltransferase (SHMT) to identify regions of conformational flexibility that could influence its mutational tolerance and adaptability [40,41]. Six complementary predictors—PONDR® FIT, PONDR® VSL2, PONDR® VL3, PONDR® VLXT, IUPred Short, and IUPred Long—were used to generate a mean disorder profile (MDP) for each residue. This consensus approach reduces predictor-specific bias and enhances reliability. Residues with scores above 0.5 were classified as intrinsically disordered, while those between 0.15 and 0.5 were considered flexible but ordered [42].

AFflecto is a web server (https://moma.laas.fr/applications/AFflecto/) designed for generating large conformational ensembles of proteins that include both structured domains and intrinsically disordered regions (IDRs), starting from structural models predicted by AlphaFold. It analyzes the structural properties of the AlphaFold model using per-residue confidence scores (pLDDT) to identify IDRs based on structural context and classifies them as tails, linkers, or loops. To explore conformational diversity within these flexible regions, AFflecto employs computationally efficient stochastic sampling algorithms . Additionally, it incorporates methods to identify conditionally folded IDRs that AlphaFold may incorrectly predict as folded elements, thereby correcting for potential misclassifications (PMID: 40133775). Through this approach, AFflecto generates protein ensembles that provide a realistic representation of structural heterogeneity.[41]

*Graph-Based Analysis*

Following mutual information (MI) calculations, we conducted a graph-based analysis to explore co-evolving residues in *E. coli* SHMT [33][34][35]. In this analysis, residues were represented as nodes, and edges indicated strong MI-based connections between them. The strength of interactions was categorized using MI thresholds: strong (MI > 0.8), moderate (0.6 < MI < 0.8), and weak (0.4 < MI < 0.6) [31][32].

This network representation provided insights into how mutations at specific residues could influence the protein's structural stability and potential drug resistance, highlighting interdependencies between residues [50][28]. To identify tightly connected sub-networks, we applied Maximal Clique Analysis (MCA) to detect fully interconnected residue groups, which often correspond to structurally or functionally dependent regions [33]. Additionally, community detection was used to partition the residue interaction network into clusters of residues more likely to interact within the same community than with residues outside their community [39].

These combined approaches provided a deeper understanding of the co-evolutionary relationships within SHMT and revealed regions critical for maintaining protein stability and mediating resistance potential [28][50].

*Fitness Trajectory Model*

To investigate the adaptive evolution of *E. coli* serine hydroxymethyltransferase (SHMT), we developed a hybrid probabilistic–deterministic computational framework that integrates residue-level fitness modeling with three-dimensional adaptive trajectory simulations. This framework captures both stochastic exploratory drift and deterministic fitness ascent, allowing quantitative visualization of how sequential mutations traverse the rugged topography of the SHMT fitness landscape [59–62]. Model parameters were tuned using experimentally derived epistatic fitness data from deep mutational scanning, enabling consistent representation of beneficial, neutral, and deleterious substitutions [11, 54].

Each single amino acid substitution was represented as a node in a weighted fitness graph, with edges denoting possible mutational transitions. The fitness value of each mutation, $f(m_i)$, quantified its relative effect on enzyme function. The fitness distance between two mutations $m_i$and $m_j$was defined as the absolute difference between their normalized fitness scores:

$$d(m_i, m_j) = | f(m_i) - f(m_j) |$$

An evolutionary path, $P = \{m_1, m_2, \dots, m_k\}$, was defined as the sequence of mutational steps that minimized the cumulative fitness distance:

$$P = arg\ _P \sum_{i=1}^{k-1} d(m_i, m_{i+1})$$

Transitions were classified as neutral when the fitness change between consecutive mutations was below a biologically negligible threshold $\varepsilon$:

$$| f(m_j) - f(m_i) | < \varepsilon$$

where $\varepsilon$typically ranged between 0.01 and 0.02, representing nearly neutral or compensatory substitutions that allow traversal across flat fitness plateaus or epistatic saddles [55, 56, 61]. Each residue-specific adaptive surface was modeled as a continuous two-dimensional Gaussian field composed of multiple fitness basins representing local optima and suboptimal intermediates. The total fitness $F(x, y)$at any coordinate was defined as the superposition of $n$Gaussian functions:

$$F(x, y) = \sum_{i=1}^{n} A_i\, exp \exp\left[ -\frac{(x - \mu_{x,i})^2 + (y - \mu_{y,i})^2}{2\sigma_i^2} \right]$$

where

- $A_i$is the amplitude (height) corresponding to the local adaptive peak,
- $(\mu_{x,i}, \mu_{y,i})$are the coordinates of the fitness maxima,
- $\sigma_i$denotes the spread (ruggedness) of each basin, and
- $n$represents the number of adaptive peaks contributing to the surface.

The surfaces were normalized such that $F(x, y) \in [0,1]$, ensuring comparability between residues. Additional shallow depressions were introduced to emulate epistatic traps and local minima, reproducing the ruggedness typical of biological fitness landscapes [59, 60].

The local fitness gradient field, capturing the direction and steepness of adaptive ascent, was computed as:

$$\nabla F(x,y) = \left(\frac{\partial F}{\partial x}, \frac{\partial F}{\partial y}\right)$$

and its magnitude $|\nabla F|$ defined the local slope intensity across the landscape.

Adaptive evolution was simulated as discrete adaptive walks over the continuous Gaussian fitness surface. Each trajectory began from the wild-type (WT) coordinate $(x_0, y_0)$ and proceeded through a sequence of points $P = \{(x_0, y_0), (x_1, y_1), \ldots, (x_k, y_k)\}$, where each successive step corresponded to a potential mutational transition.

The next position was selected according to a hybrid exploration–exploitation rule:

$$(x_{t+1}, y_{t+1}) = arg \max_{(x,y) \in N_t} [F(x,y) + \lambda\, \xi(x,y)]$$

where

- $N_t$ denotes the neighborhood of accessible mutations around the current coordinate $(x_t, y_t)$,
- $\xi(x,y) \sim N(0, g_t^2)$ introduces Gaussian stochastic noise to simulate exploratory drift,
- $g_t$ defines the exploration temperature controlling randomness, and
- $\lambda$ scales the stochastic term relative to deterministic fitness ascent.

At early stages, the model favors exploration, enabling traversal through neutral or mildly deleterious zones; as the gradient magnitude increases beyond a critical threshold $g_{crit}$, the system transitions to exploitation, following deterministic steepest ascent:

$$(x_{t+1}, y_{t+1}) = arg \max_{(x,y) \in N_t} F(x,y)$$

The change in fitness between successive steps was expressed as:

$$\Delta F_t = F(x_{t+1}, y_{t+1}) - F(x_t, y_t)$$

Transitions with $|\Delta F_t| < \varepsilon$ were designated as neutral steps, reflecting quasi-stationary or epistatic zones between adaptive peaks .

Model parameters were empirically optimized as: exploration probability $p_e = 0.7$, maximum exploration steps $n_e = 5$, and gradient threshold $g_t = 0.1$, ensuring a balanced transition between stochastic search and deterministic optimization .

Each simulated trajectory was rendered on its respective Gaussian fitness surface using Matplotlib 3D visualization (plot_surface), with the *z*-axis representing normalized fitness and color-coded using the viridis colormap. The adaptive walk was overlaid as a smooth 3D curve (plot3D), constrained to the curvature of the surface to accurately depict the evolutionary route.

Key trajectory states were annotated as:

- Blue sphere: Wild-type starting state (WT)
- Orange points: Intermediate or transitional mutations
- Gray markers: Neutral drift transitions ($|\Delta F_t| < \varepsilon$)
- Yellow sphere: Final adaptive peak (fitness maximum)

All trajectories were plotted in a continuous 3D frame to illustrate the transition from WT to the global or local optimum. Steep segments corresponded to deterministic gradient ascent, while flat or oscillatory segments indicated neutral drift and epistatic buffering.

Residues exhibiting smooth Gaussian slopes reflected robust adaptive channels conducive to continuous improvement, whereas rugged multi-peaked surfaces represented constrained evolvability and frequent entrapment in local optima. Extending this modeling across multiple residues (e.g., S35, Y202, D235) enabled comparative quantification of adaptive path length, fitness gain, and neutral drift frequency, thereby delineating site-specific differences in the evolutionary plasticity of SHMT.

## SUPPLEMENTARY FIGURES

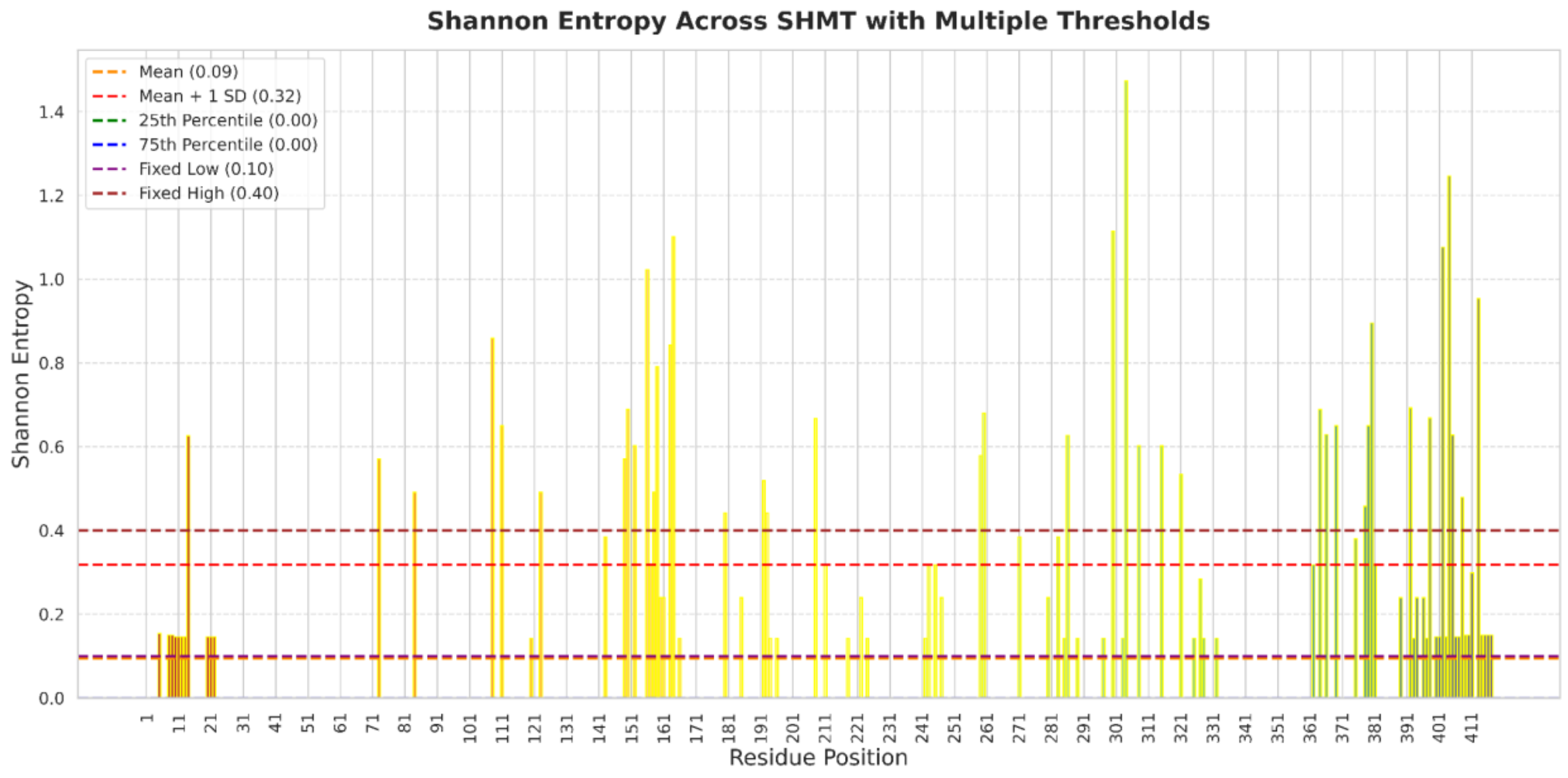

**Supplementary Figure 1.** Shannon entropy values are plotted per residue position throughout the SHMT protein sequence; higher values indicate sequence variability, while lower values indicate conserved residues. The mean entropy (orange dashed line), mean + 1 standard deviation (red dashed line), 25th percentile (green dashed line), 75th percentile (blue dashed line), and fixed thresholds at 0.10 (purple dashed line) and 0.40 (brown dashed line) are the six thresholding criteria used to classify

residue variability that are superimposed on the bar plot. Strong comparative and functional analyses of SHMT across various homologues are supported by these complementary thresholds, which enable the systematic identification of conserved and variable sites. conservation probably because of its structural or functional significance. Conversely, high entropy positions such as 156, 308, and 399 indicate evolutionary flexibility or surface

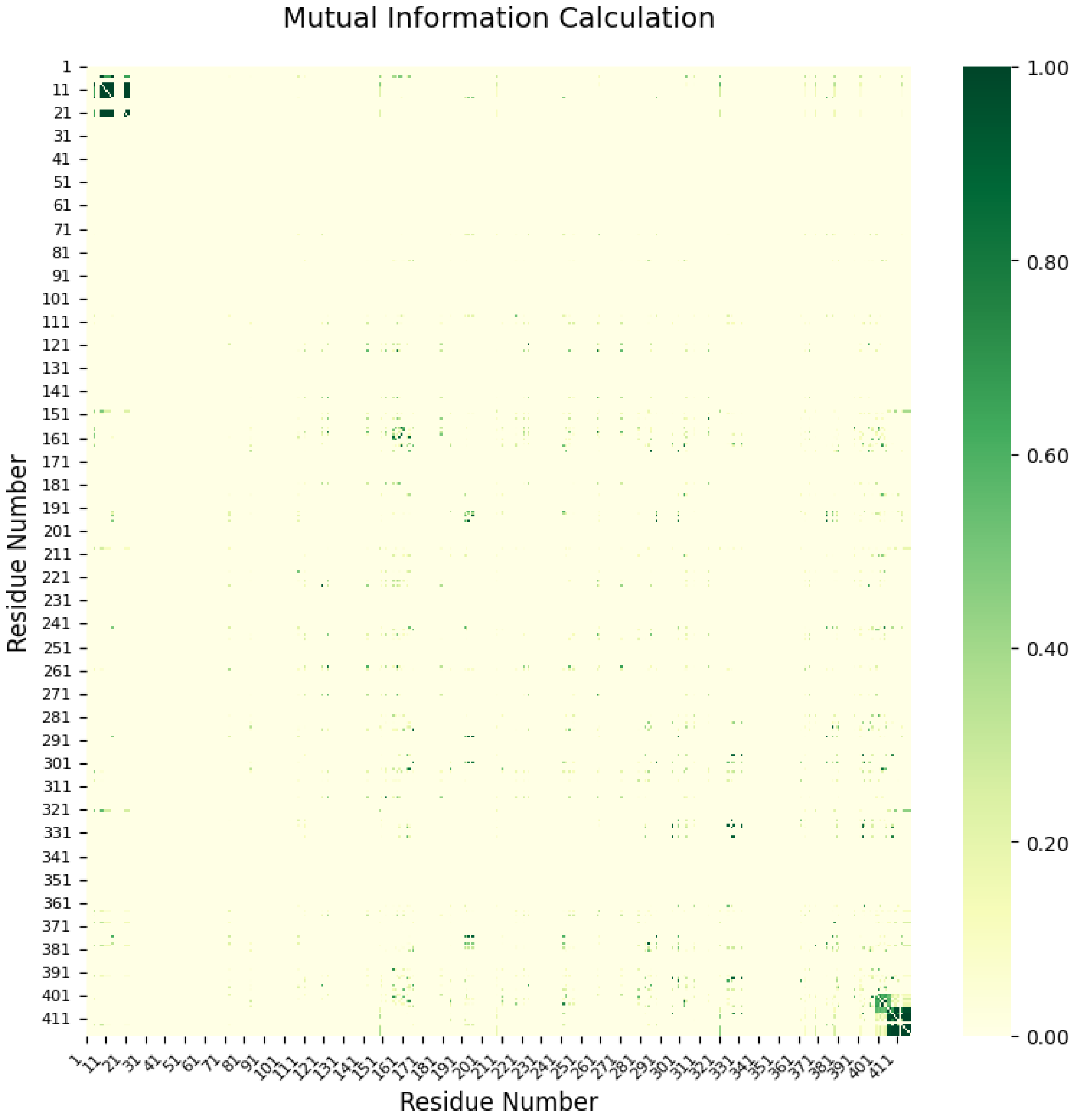

**Supplementary Figure 2.** Mutual Information (MI) matrix representing residue-residue co-variation across the SHMT multiple sequence alignment. Each pixel in the heatmap corresponds to an MI score between a pair of residue positions, with values ranging from 0 (no mutual dependence) to 1 (maximum co-variation), as indicated by the colour bar. Strong MI signals are observed between several residue

pairs in the N-terminal (e.g., residues 1–20) and C-terminal (e.g., residues 390–410) regions, suggesting potential evolutionary or structural coupling. Scattered moderate MI values across the central matrix may indicate long-range dependencies or allosteric relationships

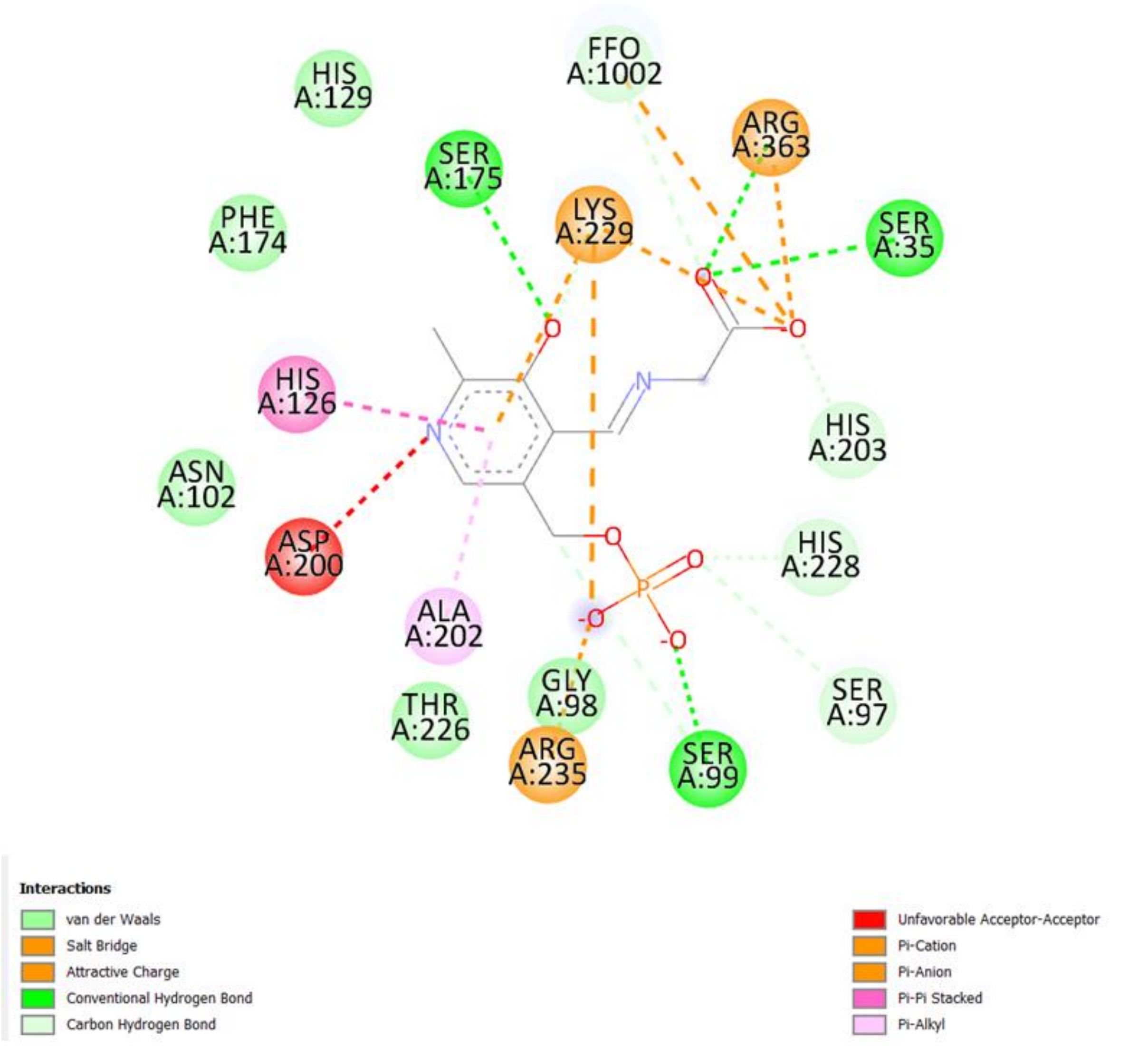

**Supplementary Figure 3.** Detailed ligand interaction map depicting the noncovalent molecular interactions between the SHMT enzyme and the bound ligand FFO (folate analogue). The central hexagonal ring represents the ligand, while surrounding residues are coloured and labelled according to interaction type and chain position. A diverse array of molecular forces stabilises ligand binding within the active site. Electrostatic interactions such as salt bridges (orange dashed lines) are formed notably between LYS229 and ARG235 with the ligand, suggesting a strong role in anchoring the molecule within the binding pocket. Conventional hydrogen bonds (green dashed lines) are observed with residues such as SER99, GLY98, and HIS203, contributing to ligand orientation and specificity. Carbon-hydrogen bonds (light green) and attractive charge interactions (solid orange) add further stabilisation. Pi-related interactions, including pi-pi stacking (pink dashed lines) with HIS126 and pi-cation interactions (orange solid lines) with LYS229 and ARG363, highlight the aromatic character of the binding environment. An unfavourable donor-acceptor interaction (solid red) is noted between ASP200 and the ligand, possibly reflecting a strained or transient configuration

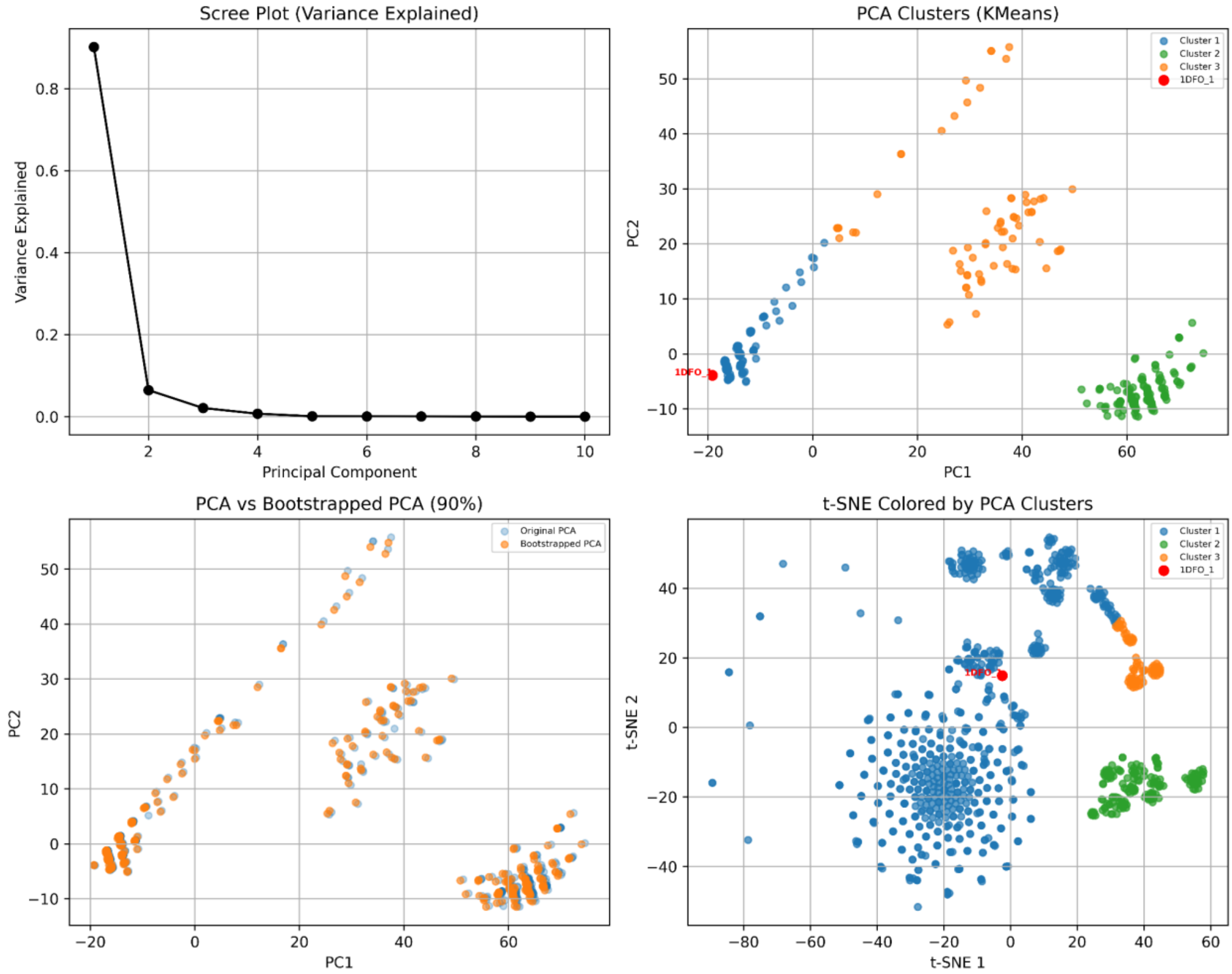

**Supplementary Figure 4:** Multivariate analysis of SHMT sequence divergence using dimensionality reduction and clustering methods. (A) The scree plot shows the proportion of variance explained by the first 10 principal components derived from the standardised Hamming distance matrix of SHMT homologs, with PC1 accounting for the largest share. (B) PCA projection of sequences onto the first two components reveals three distinct KMeans clusters (k=3), with the E. coli reference sequence “1DFO” highlighted in red. (C) Comparison of the original PCA with a bootstrapped PCA based on 90% random sampling demonstrates high overlap, confirming robustness of the dimensionality reduction. (D) t-SNE visualisation based on the same dataset shows clustering patterns consistent with PCA while emphasising local sequence similarities; the reference sequence "1DFO" is again distinctly marked.

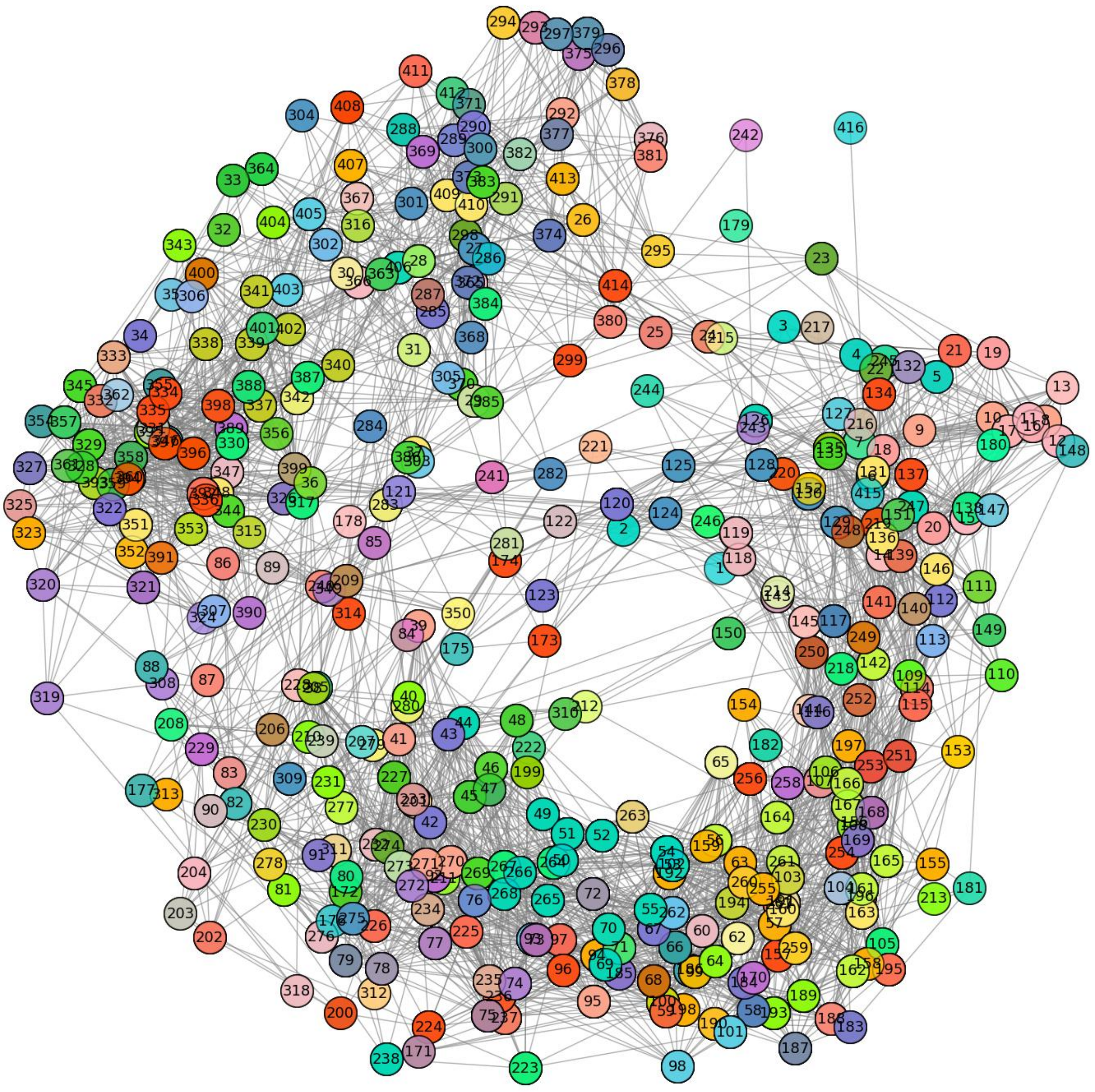

**Supplementary Figure 5. Residue-level DCCM-based maximal clique network of *E. coli* SHMT.** The network was constructed from the Dynamic Cross-Correlation Matrix (DCCM) of backbone atomic fluctuations. Each node represents a single amino acid residue, numbered according to the SHMT sequence. Edges indicate statistically significant correlated motions (correlation coefficient > threshold), capturing long- and short-range dynamic couplings. Maximal cliques subsets of residues in which every residue is directly connected to every other member were identified to delineate highly

cooperative structural units. Distinct colors denote individual cliques, highlighting spatially contiguous and dynamically coordinated residue clusters across the protein. This residue-level partitioning reveals how SHMT dynamics are organized into localized cooperative groups that may underpin catalytic activity, structural stabilization, and allosteric communication.

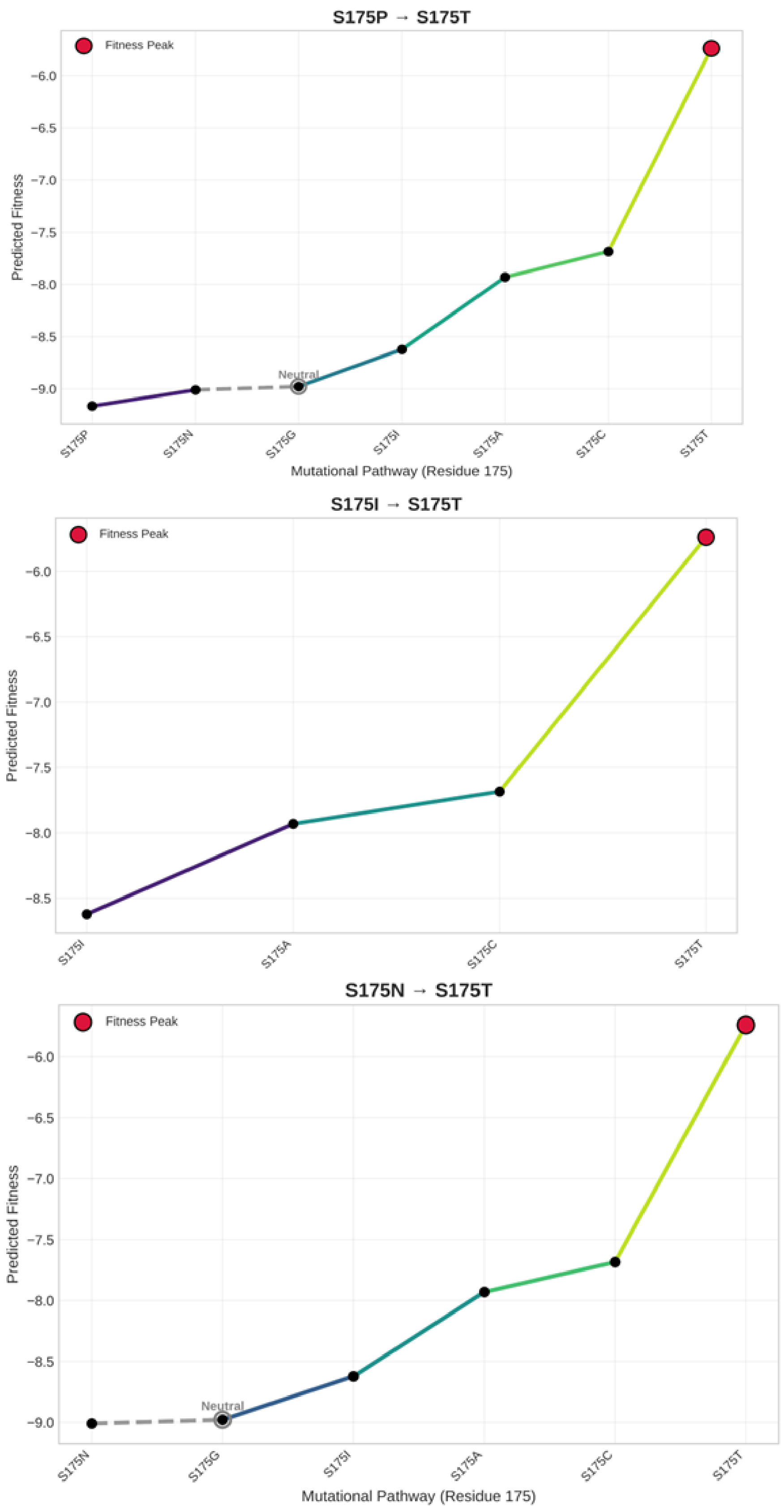

**Supplementary Figure 6. Multi-step adaptive walks converging to the S175T fitness peak in *E. coli* SHMT:** Predicted fitness trajectories are shown for three distinct starting variants—S175P, S175I,

and S175N—each evolving toward the adaptive peak represented by S175T. The y-axis represents predicted fitness values derived from the mutational landscape model, while the x-axis indicates sequential mutational intermediates along the residue 175 pathway. Solid colored lines trace favorable adaptive steps, with gradient hues corresponding to fitness gains, whereas dashed grey lines denote neutral or nearly neutral transitions. Red-highlighted nodes indicate the fitness peak (S175T), marking convergence across independent trajectories. Collectively, these plots reveal parallel routes toward a shared adaptive optimum despite differing mutational starting points, illustrating constrained but convergent evolvability at this catalytic residue.

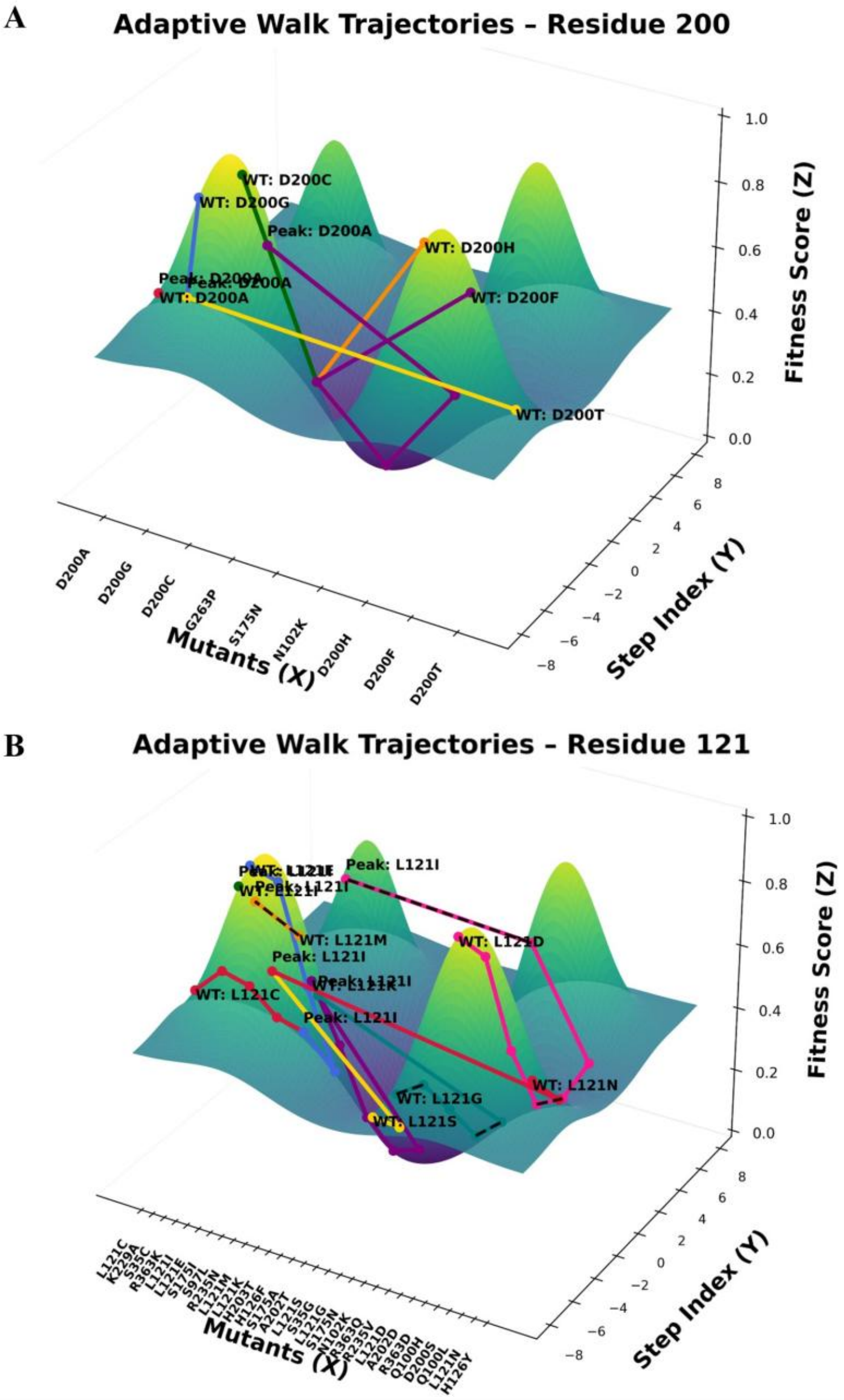

**Supplementary Figure 7. Adaptive Walk Trajectories for SHMT Residues 200 and 121 Derived From Epistatic Fitness Landscapes:** This figure shows how mutations at residues 200 (panel A) and 121 (panel B) move across the predicted fitness landscape during simulated adaptive walks. The x-axis

represents the different mutant states, while the step index tracks the order in which a trajectory progresses, and the z-axis shows the relative fitness score. For both residues, paths often move toward local peaks, but they do so through different mutational routes and with varying degrees of constraint. Residue 200 displays smoother, more directed trajectories, where several variants (such as D200A, D200C, and D200H) sit near prominent fitness peaks. In contrast, residue 121 shows a more rugged landscape with multiple competing peaks and several alternative routes, reflecting a higher level of mutational flexibility. Together, these trajectories highlight how even within the same active-site region, different residues can follow distinct evolutionary pathways depending on how tolerant or restrictive their local structural environment is.

**SUPPLEMENTARY TABLE 1**

| **Community** | **Residues** | Macro-community |
|---|---|---|
| Community 1 (39 residues) | 106, 109, 257, 238, 199, 108, 100, 222, 93, 172, 258, 256, 254, 235, 107, 236, 259, 94, 255, 104, 224, 260, 96, 103, 92, 101, 95, 105, 97, 99, 91, 225, 98, 201, 198, 223, 237, 171, 102 | 2 |
| Community 2 (16 residues) | 208, 203, 200, 206, 202, 209, 213, 229, 230, 204, 212, 210, 211, 207, 228, 205 | 2 |
| Community 3 (12 residues) | 252, 249, 251, 243, 246, 247, 244, 250, 253, 245, 248, 242 | 2 |
| Community 4 (9 residues) | 220, 218, 216, 221, 197, 214, 217, 219, 215 | 1 |
| Community 5 (38 residues) | 273, 231, 233, 45, 77, 234, 41, 44, 276, 226, 43, 267, 232, 274, 269, 277, 227, 38, 76, 270, 50, 46, 51, 49, 275, 40, 278, 265, 47, 42, 78, 268, 271, 272, 264, 266, 39, 48 | 2 |
| Community 6 (37 residues) | 354, 348, 326, 325, 324, 359, 308, 330, 345, 321, 329, 346, 323, 349, 361, 332, 331, 333, 328, 352, 350, 356, 360, 351, 309, 327, 355, 362, 320, 357, 322, 344, 121, 353, 347, 319, 358 | 2 |
| Community 7 (10 residues) | 307, 301, 305, 303, 306, 302, 299, 298, 300, 304 | 2 |
| Community 8 (12 residues) | 315, 311, 314, 178, 313, 312, 179, 177, 316, 180, 318, 317 | 2 |
| Community 9 (11 residues) | 280, 282, 283, 286, 281, 289, 285, 288, 279, 287, 284 | 3 |
| Community 10 (27 residues) | 71, 262, 70, 66, 56, 58, 64, 60, 63, 261, 54, 61, 72, 68, 67, 263, 62, 73, 65, 74, 59, 69, 57, 53, 52, 75, 55 | 1 |
| Community 11 (40 residues) | 173, 114, 142, 174, 132, 110, 136, 119, 111, 137, 116, 145, 123, 122, 130, 113, 133, 118, 143, 115, 125, 126, 129, 144, 134, 120, 124, 140, 127, 117, 135, 139, 138, 176, 141, 112, 128, 131, 175, 170 | 2 |
| Community 12 (52 residues) | 374, 365, 292, 24, 26, 23, 363, 414, 29, 373, 364, 36, 31, 377, 372, 412, 28, 371, 415, 368, 413, 343, 366, 416, 376, 30, 370, 291, 367, 369, 27, 33, 294, 411, 34, 342, 290, 408, 375, 35, 407, 409, 293, 295, 297, 32, 25, 37, 296, 410, 405, 406 | 4 |
| Community 13 (22 residues) | 8, 21, 16, 22, 18, 12, 5, 4, 6, 3, 19, 9, 13, 7, 1, 20, 15, 10, 17, 14, 11, 2 | 1 |
| Community 14 (35 residues) | 384, 336, 334, 398, 393, 341, 385, 382, 335, 388, 383, 391, 380, 387, 340, 389, 401, 337, 402, 397, 392, 386, 396, 339, 394, 338, 399, 395, 379, 390, 403, 400, 378, 381, 404 | 1 |
| Community 15 (32 residues) | 168, 164, 156, 188, 160, 184, 162, 167, 195, 181, 154, 159, 169, 163, 155, 191, 182, 166, 186, 157, 158, 193, 161, 165, 189, 183, 194, 192, 187, 196, 185, 190 | 2 |
| Community 16 (15 residues) | 89, 88, 86, 239, 82, 85, 80, 240, 79, 83, 87, 241, 90, 84, 81 | 1 |
| Community 17 (9 residues) | 152, 148, 146, 310, 150, 153, 149, 147, 151 | 2 |

**SUPPLEMENTARY TABLE 1**: Community and Macro-Community Assignment of Residues in SHMT Network Analysis. The table lists the identified residue communities derived from structure

network analysis of SHMT. Each community consists of a group of interacting residues (shown by residue numbers) and is associated with its corresponding macro-community classification.

**Declarations of interest:** The authors do not have any conflicts of interest.

**Funding and Acknowledgment:** DP and SC acknowledge the Director, Birla Institute of Technology and Science-Pilani, and Head, Department of Biological Sciences, BITS-Pilani Hyderabad, India. The work is supported by BITS NFSG (N4/24/1032). DP acknowledges Birla Institute of Technology and Science-Pilani Hyderabad for doctoral fellowship.

**Author Contribution:** S.C. D.S and D.P. conceptualized the overall project outline. D.P and S.C designed the work plans. D.P performed the computational calculations. D.P and D.S performed the analyses and prepared figures. D.P wrote the paper. V.U performed the disorder calculation. D.Z, D.S, S.C and V.U reviewed the manuscript.

**Data Availability Statement:** All data supporting the findings of this study are included within the manuscript and its supplementary materials.